\documentclass[twocolumn,tighten,times]{aastex63} 

\usepackage{amsmath} 
\usepackage{amssymb}
\extrafloats{100}

\received{\today}
\revised{\today}
\accepted{\today}

\newcommand{\galprop}{\textsc{GalProp}}
\newcommand{\helmod}{\textsc{HelMod}}

\shorttitle{
Spectra of He isotopes and the $^3$He/$^4$He ratio}
\shortauthors{Boschini et al.}

\begin{document}

\title{
Spectra of He isotopes and the $^3$He/$^4$He ratio
}

\author[0000-0002-6401-0457]{M.~J.~Boschini}
\affiliation{INFN, Milano-Bicocca, Milano, Italy}
\affiliation{CINECA, Segrate, Milano, Italy}

\author[0000-0002-7669-0859]{S.~{Della~Torre}}
\affiliation{INFN, Milano-Bicocca, Milano, Italy}

\author[0000-0003-3884-0905]{M.~Gervasi}
\affiliation{INFN, Milano-Bicocca, Milano, Italy}
\affiliation{Physics Department, University of Milano-Bicocca, Milano, Italy}

\author[0000-0003-1942-8587]{D.~Grandi}
\affiliation{INFN, Milano-Bicocca, Milano, Italy}
\affiliation{Physics Department, University of Milano-Bicocca, Milano, Italy}

\author[0000-0003-1458-7036]{G.~J\'{o}hannesson} 
\affiliation{Science Institute, University of Iceland, Dunhaga 3, IS-107 Reykjavik, Iceland}

\author[0000-0002-2168-9447]{G.~{La~Vacca}}
\affiliation{INFN, Milano-Bicocca, Milano, Italy}
\affiliation{Physics Department, University of Milano-Bicocca, Milano, Italy}

\author[0000-0002-3729-7608]{N.~Masi}
\affiliation{INFN, Bologna, Italy}
\affiliation{Physics Department, University of Bologna, Bologna, Italy}

\author[0000-0001-6141-458X]{I.~V.~Moskalenko} 
\affiliation{Hansen Experimental Physics Laboratory, Stanford University, Stanford, CA 94305}
\affiliation{Kavli Institute for Particle Astrophysics and Cosmology, Stanford University, Stanford, CA 94305}

\author{S.~Pensotti}
\affiliation{INFN, Milano-Bicocca, Milano, Italy}
\affiliation{Physics Department, University of Milano-Bicocca, Milano, Italy}

\author[0000-0002-2621-4440]{T.~A.~Porter} 
\affiliation{Hansen Experimental Physics Laboratory, Stanford University, Stanford, CA 94305}
\affiliation{Kavli Institute for Particle Astrophysics and Cosmology, Stanford University, Stanford, CA 94305}

\author{L.~Quadrani}
\affiliation{INFN, Bologna, Italy}
\affiliation{Physics Department, University of Bologna, Bologna, Italy}

\author[0000-0002-1990-4283]{P.~G.~Rancoita}
\affiliation{INFN, Milano-Bicocca, Milano, Italy}

\author[0000-0002-7378-6353]{D.~Rozza}
\affiliation{INFN, Milano-Bicocca, Milano, Italy}

\author[0000-0002-9344-6305]{M.~Tacconi}
\affiliation{INFN, Milano-Bicocca, Milano, Italy}
\affiliation{Physics Department, University of Milano-Bicocca, Milano, Italy}




\begin{abstract}

Since its launch, the Alpha Magnetic Spectrometer--02 (AMS-02) has delivered outstanding quality measurements of the spectra of cosmic-ray (CR) species, $e^{\pm}$, $\bar{p}$, and nuclei (H--Si, S, Fe), which have resulted in a number of breakthroughs. Besides elemental spectra, AMS-02 also measures the spectra of light isotopes albeit within a smaller rigidity range. In this paper, we use the precise measurements of He isotopes and the $^3$He/$^4$He ratio by AMS-02, together with Voyager 1 data, and investigate their origin. We show that there is an excess in $^3$He at higher energies compared to standard calculations assuming its fully secondary origin, which may indicate a source of primary $^3$He, or signify a change in the propagation parameters over the Galactic volume probed by the $^3$He/$^4$He ratio. We provide updated local interstellar spectra (LIS) of $^3$He and $^4$He in the rigidity range from 2--40 GV. Our calculations employ the self-consistent \galprop{}--\helmod{} framework that has proved to be a reliable tool in deriving the LIS of CR $e^{-}$, $\bar{p}$, and nuclei $Z\le28$.
\end{abstract}



\section{Introduction} \label{Intro}

Precise data at low and high energies delivered by modern space instrumentation have triggered a number of discoveries of new features in the spectra of CR species. The combined data from individual experiments covers an enormous range of rigidities, from few MV to tens of TV, where the individual spectra of CR species are shaped by many different processes. These data enable the probing of the stellar nucleosynthesis, properties of the interstellar medium (ISM), and the origin of CRs.

The AMS-02 collaboration has now completed a series of papers on CR species H--Si, S, and Fe \citep{2021PhRvL.126d1104A, 2021PhRvL.126h1102A, 2021PhRvL.127b1101A, 2021PhR...894....1A, 2023PhRvL.130u1002A}. Besides the elemental spectra, AMS-02 also measures the spectra of light isotopes, albeit in a smaller rigidity range. In this paper, we use precise measurements of He isotopes and the $^3$He/$^4$He ratio by AMS-02 \citep{2019PhRvL.123r1102A}, together with Voyager 1 data, to test their origin. We also provide new and updated $^3$He and $^4$He LIS in the rigidity range from 2--40 GV. Our calculations and interpretation employ the \galprop{}\footnote{Available from http://galprop.stanford.edu \label{galprop-site}}--\helmod{}\footnote{https://www.helmod.org/; the website operates on a virtual server at the Italian Space Agency, ASI.  \label{helmod-footnote}} framework that has proven to be a reliable tool for modeling the LIS of CR species \citep{2019HelMod, 2020ApJS..250...27B}.

\section{Calculations} \label{calcs}

In this work we employ the same CR propagation model phenomenology with distributed reacceleration and convection from our previous analyses \citep[for more details see][]{2017ApJ...840..115B,  2018ApJ...854...94B, 2018ApJ...858...61B, 2020ApJS..250...27B, 2020ApJ...889..167B, 2021ApJ...913....5B, 2022ApJ...925..108B}, but use the latest version 57.1 of the \galprop{} code for Galactic propagation of CRs and diffuse emissions as described in detail by \citet{2022ApJS..262...30P}, see also \citet{2020ApJS..250...27B}, and references therein. This \galprop{} release also includes new and improved parameterizations for calculations of the total inelastic cross sections for $p+A$ and He $+ A$ reactions, together with new routines for the production cross sections for isotopes of hydrogen $^{2,3}$H and helium $^{3,4}$He in $p+A$ and He $+ A$ reactions, as well as $^2$H production in the $pp$-fusion reaction. Therefore, the accuracy of calculations of production and propagation of isotopes of helium is significantly improved compared to earlier \galprop{} releases. These cross sections are the best available so far.

Full details of the \helmod{} code version 4 for heliospheric propagation are provided by \citet{2019AdSpR..64.2459B}. It solves the Fokker-Planck equation for heliospheric propagation in Kolmogorov formulation backward in time \citep{2016JGRA..121.3920B}. The last version of the code was presented in \citet{BOSCHINI2024} where the computational algorithm is re-designed to work on GPU architecture.
Moreover, in \citet{BOSCHINI2024}, \helmod{} parameters are revisited using the updated daily proton flux measured by AMS-02 \citep{2021PhR...894....1A} and focusing on the descending phase of solar cycle 24, i.e. from $\sim$2015--2020. The updates are motivated by the latest high-precision measurements from the AMS-02 detector which revealed for the first time the fine features of GCR fluxes' evolution during a period of positive interplanetary magnetic field polarity. The accuracy of the solution was tested in \citet{2016JGRA..121.3920B} using the Crank-Nicholson technique and found to be better than 0.5\% at low rigidities. In \citet{BOSCHINI2024} we demonstrate that the uncertainties of the numerical solution decrease with increasing the energy and are lower than the equivalent Poisson uncertainties evaluated using the number of simulated Monte Carlo events. The large number of simulated events ($>$5000) ensures that the statistical errors are negligible compared to the other modeling uncertainties.

When comparing our calculations with data collected over an extended period of time, variations in the solar activity are addressed in the following way. The propagation equation is solved for each Carrington rotation, and the numerical results are then combined with the instrument exposure over the appropriate time period. This approach is equivalent to using a weighted average that accounts for both exposure time and absolute counting rate variations.

The CR propagation model parameters in the ISM, along with their confidence limits, are derived from the best available CR data for all species H$-$Ni, and all rigidities from a few MV to TV range using a Markov Chain Monte Carlo (MCMC) routine. We use the same method as described by \citet{2017ApJ...840..115B}. The five main propagation parameters that affect the overall shape of CR spectra were left free in the scan, using \galprop{} running in 2D mode: the Galactic halo half-width $z_h$, the normalization of the diffusion coefficient $D_0$ at the reference rigidity $R=4$ GV and the index of its rigidity dependence $\delta$, the Alfv\'en velocity $V_{\rm Alf}$, and the gradient of the convection velocity $dV_{\rm conv}/dz$ ($V_{\rm conv}=0$ in the plane, $z=0$). Their best-fit values tuned to the AMS-02 data are listed in Table~\ref{tbl-prop} (labeled as the ``standard model'') and are the same as obtained by \citet{2020ApJS..250...27B}. The radial size of the Galaxy does not significantly affect the values of these parameters and was set to 20 kpc. We also introduced a factor $\beta^\eta$ in the diffusion coefficient, where $\beta=v/c$ is the particle velocity in units of the speed of light, and $\eta$ was left free. The latter improves the agreement at low energies, and slightly affects the choice of injection indices ${\gamma}_0$ an ${\gamma}_1$. 

\begin{figure*}[tb!]
	\centering
	\includegraphics[width=0.5\textwidth]{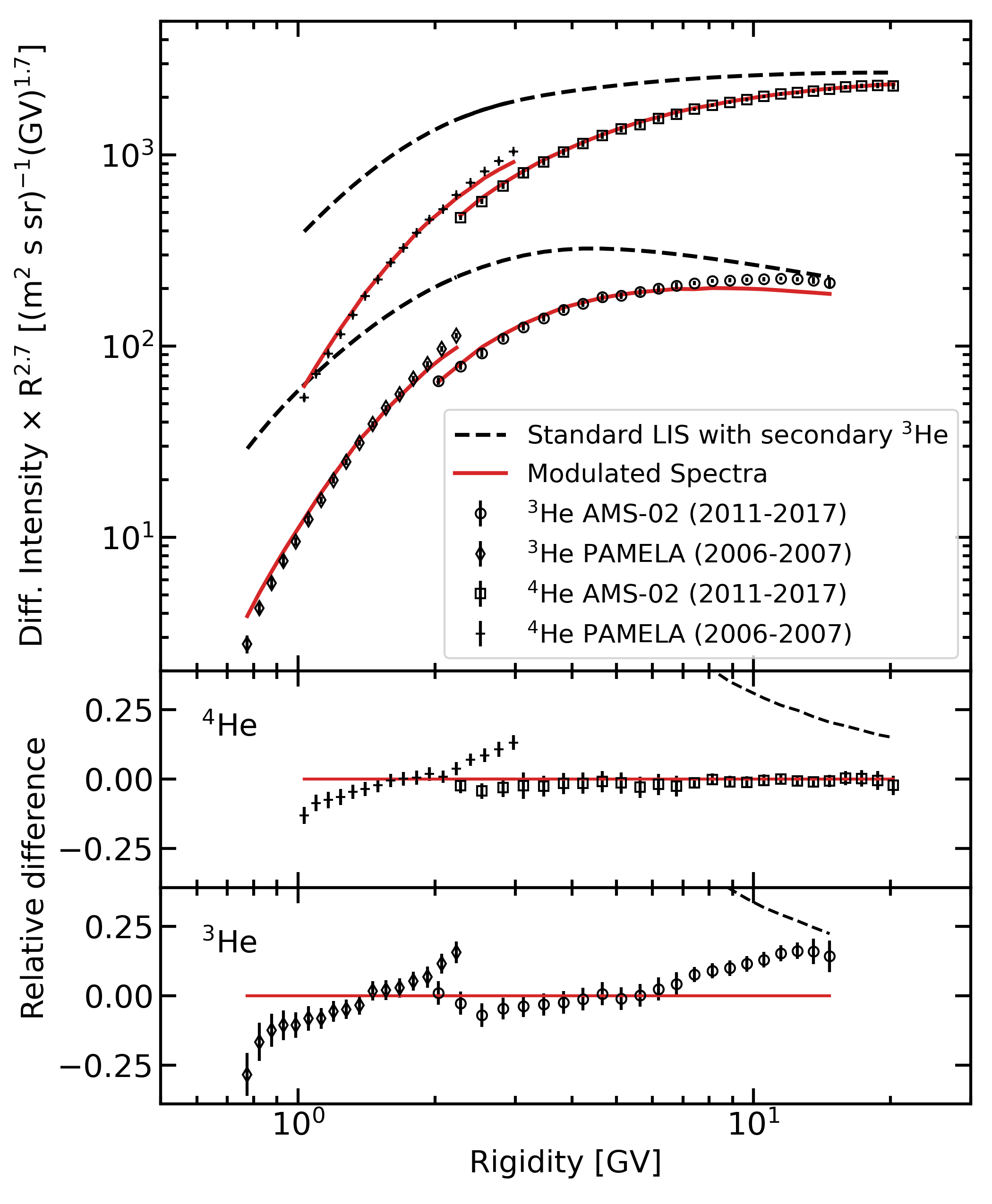}\\ 
	\includegraphics[width=0.5\textwidth]{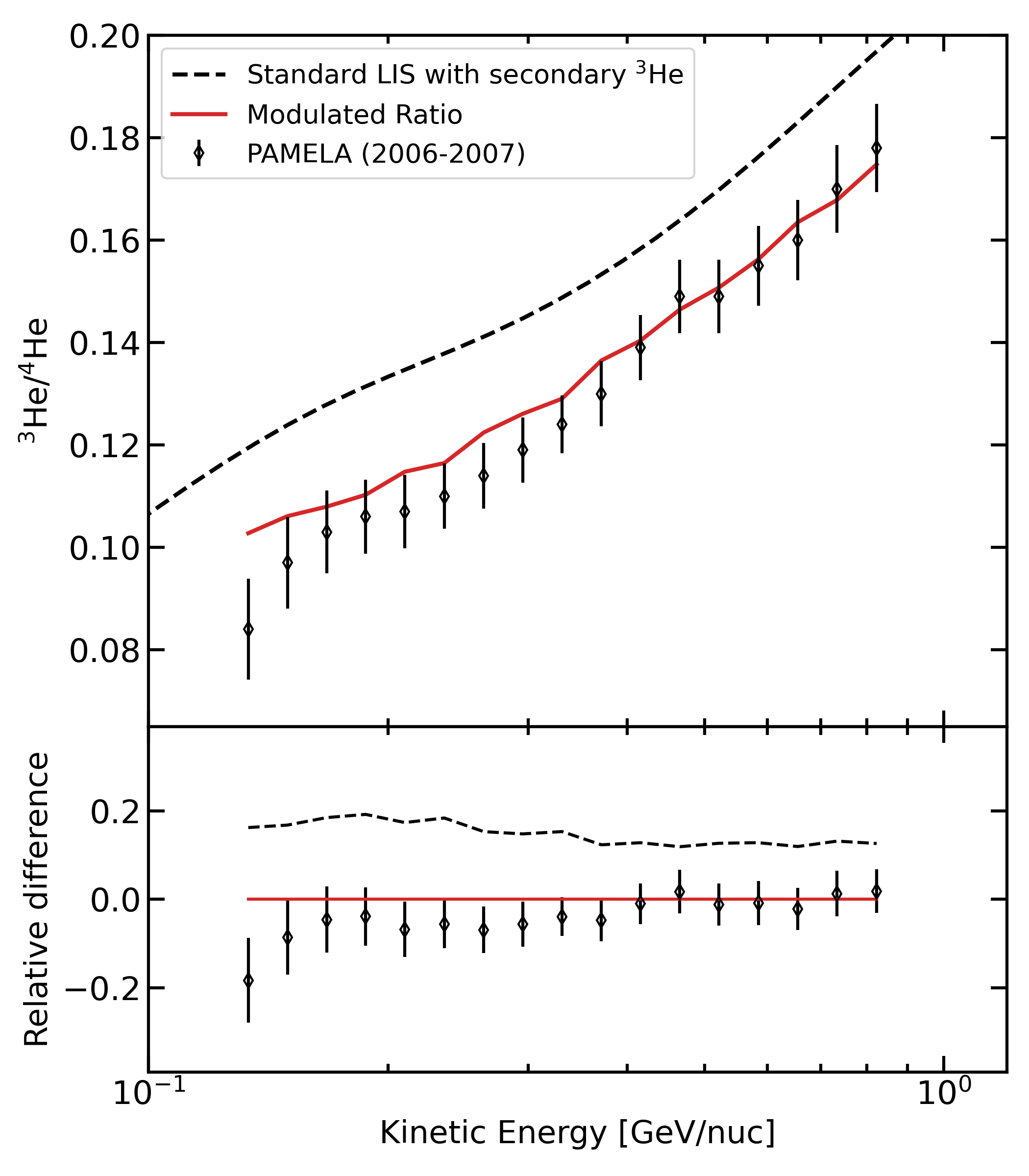}\hfill
	\includegraphics[width=0.5\textwidth]{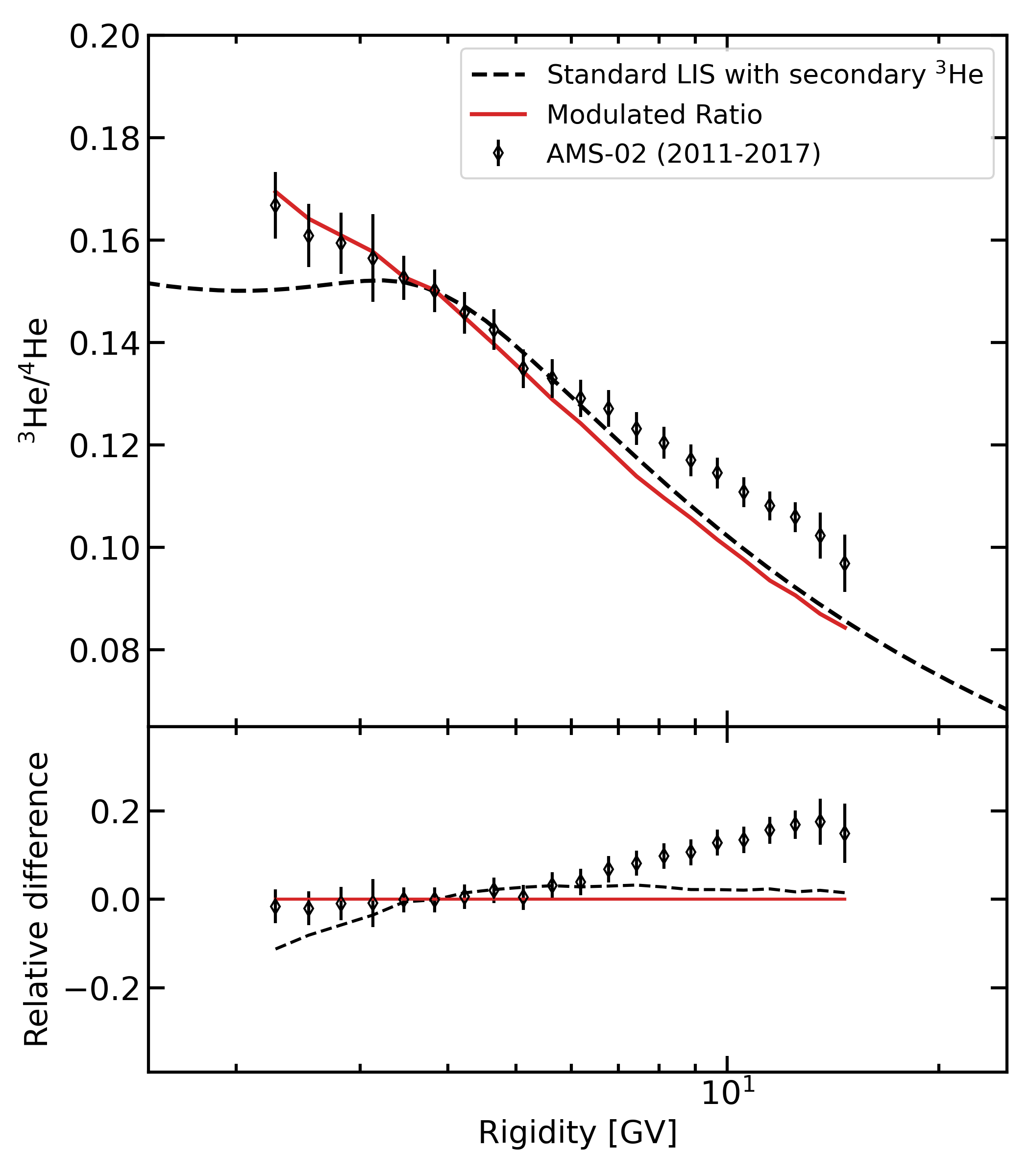} 
	\caption{	
A comparison of the standard calculations with fully secondary $^{3}$He with data. \textit{Upper panel} shows a comparison of the calculated $^{3,4}$He spectra with PAMELA \citep{2016JPhCS.675c2001M} and AMS-02 \citep{2019PhRvL.123r1102A} data. The dashed black lines show the LIS, and the solid red lines$-$the corresponding modulated spectra. \textit{Lower panels} show a comparison of the $^3$He/$^4$He ratio with PAMELA and AMS-02 data. Note different units in PAMELA and AMS-02 plots. The relative differences between our calculations and the data sets are shown in the bottom part of each panel. 
	}
	\label{fig:He-spec-default}
\end{figure*}

\begin{figure*}[tb!]
	\centering
	\includegraphics[width=0.5\textwidth]{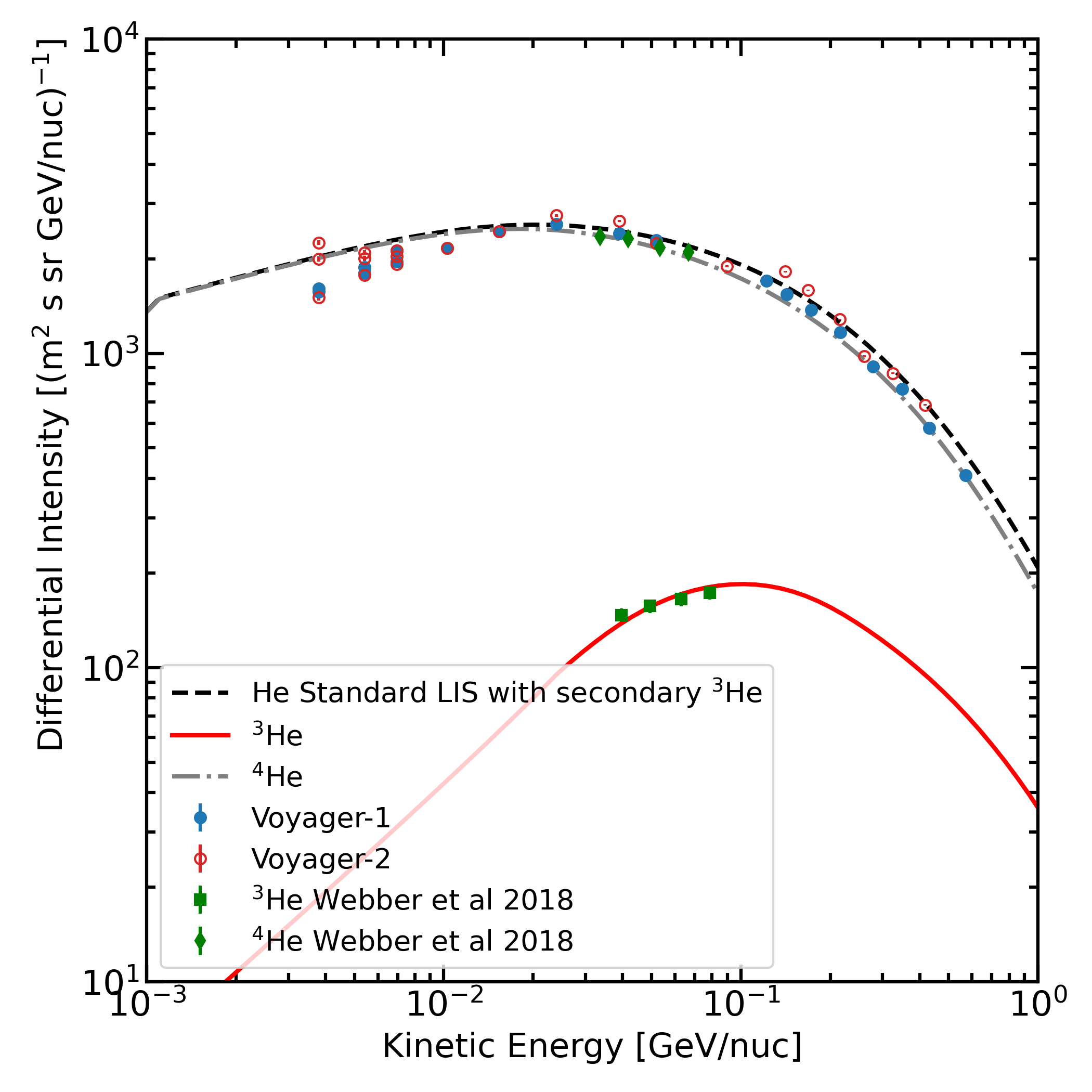}\hfill
	\includegraphics[width=0.5\textwidth]{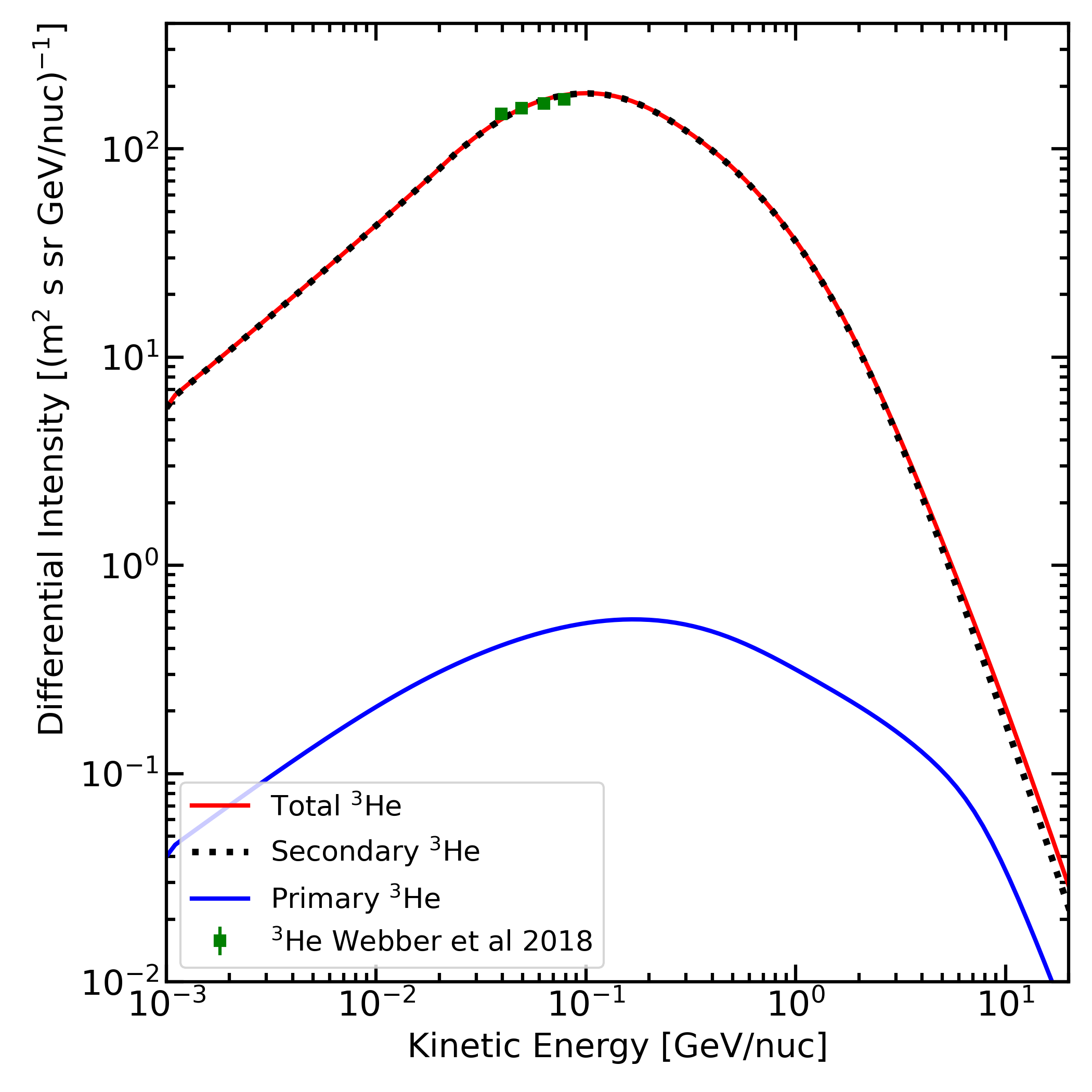} 
	\caption{	
\textit{Left panel:} A comparison of the standard calculations with fully secondary $^{3}$He with Voyager data \citep{2016ApJ...831...18C, 2018arXiv180208273W, 2019NatAs...3.1013S}. The same model as in Fig.~\ref{fig:He-spec-default}, but shows a comparison of the total He spectrum, and isotopes $^{3,4}$He with Voyager 1, 2 data. \textit{Right panel:} A model with primary $^{3}$He. The primary $^{3}$He component is shown by the blue solid line shows the, while the solid red line shows the secondary component. The thick dashed black line is the total $^{3}$He. Contribution of the primary $^{3}$He component is negligible below $\approx$5 GeV/n.
	}
	\label{fig:He-spec-default1}
\end{figure*}

The corresponding B/C ratio also remains the same \citep[see Fig.~4 of][]{2020ApJS..250...27B}, and compares well with all available measurements: Voyager~1 \citep{2016ApJ...831...18C}, ACE-CRIS\footnote{http://www.srl.caltech.edu/ACE/ASC/level2/cris\_l2desc.html} \citep{2013ApJ...770..117L}, AMS-02 \citep{2018PhRvL.120b1101A}, ATIC-2 \citep{2009BRASP..73..564P}, CREAM \citep{2008APh....30..133A, 2009ApJ...707..593A}, and NUCLEON \citep{2019AdSpR..64.2559G}. The particular choice of the production cross section parameterization has a minor effect on the derived propagation parameters given that they are all reproducing the available experimental data \citep[for more details see][]{2020ApJS..250...27B}. 

For the calculations, we use our standard configuration. The convection velocity is assumed to increase linearly with distance $z$ from the plane, $V_{\rm conv}(z)=V_0+ z\cdot dV_{\rm conv}/dz$. The spatial diffusion coefficient is parametrized as $D_{xx} = \beta^\eta D_0 R^\delta$. If reacceleration is included, the momentum-space diffusion coefficient $D_{pp}$ is related to $D_{xx}$ as $D_{pp}=p^{2}V_{\rm Alf}^{2}/\left(  9D_{xx}\right)$ \citep{1990acr..book.....B, 1994ApJ...431..705S}. See also a review by \citet{2007ARNPS..57..285S} and our papers cited in the beginning of this section. 

The injection spectra of CR species are parameterized as: 
\begin{equation}  \label{eq:1}
q(R) \propto (R/R_0)^{-\gamma_0}\prod_{i=0}^2\bigg[1 + (R/R_i)^\frac{\gamma_i - \gamma_{i+1}}{s_i}\bigg]^{s_i},
\end{equation}
where $R$ is the rigidity, $\gamma_{i =0,1,2,3}$ are the spectral indices, $R_{i = 0,1,2}$ are the break rigidities, $s_i$ are the smoothing parameters ($s_i$ is negative/positive for $|\gamma_i |\lessgtr |\gamma_{i+1} |$). The source abundance of each species is defined as the number of injected particles in the energy range from 100 MeV/n to 100 GeV/n normalized to hydrogen (H$=$$10^6$). 

\begin{deluxetable}{rlcc}[tb!]
	\def\arraystretch{0.9}
	\tablewidth{0mm}
	\tablecaption{Best-fit propagation parameters for the standard and alternative models\label{tbl-prop}} 
	\tablehead{
		&& \multicolumn{2}{c}{Model}\\
		\colhead{Parameter}& \multicolumn{1}{l}{Units}& \multicolumn{1}{c}{ Standard}& \multicolumn{1}{c}{Alternative} 
	}
	\startdata
	$z_h$ & kpc &$4.0\pm0.6$ & 4.0 (fixed)\\
	$D_0 (R= 4\ {\rm GV})$ & $10^{28}$ cm$^{2}$ s$^{-1}$  & $4.3\pm0.7$ & $5.9\pm1.0$\\ 
	$\delta$ & \nodata &$0.415\pm0.025$
								& $0.19\pm0.06$\\
	$V_{\rm Alf}$ & km s$^{-1}$ &$30\pm3$ & $27.3\pm5$\\
	$dV_{\rm conv}/dz$ & km s$^{-1}$ kpc$^{-1}$ & $9.8\pm0.8$ & $2\pm2$\\
	$\eta$ & \nodata & 0.70 & 1.2\\
	\enddata
\end{deluxetable}

\begin{deluxetable*}{ccc rcl rcl rcl rcl r}[tb!]        
	\tabletypesize{\footnotesize}
        \def\arraystretch{1.1}
        \tablecolumns{16}
        \tablewidth{0mm}
        \tablecaption{The injection spectra of primary $^{3,4}$He isotopes \label{tbl-inject}}
        \tablehead{
                \multicolumn{2}{c}{} & 
                \multicolumn{1}{c}{Source} &
                \multicolumn{10}{c}{Spectral parameters}\\
                \cline{4-16}
                \multicolumn{1}{c}{} & 
                \multicolumn{1}{c}{Model} & 
                \multicolumn{1}{c}{abundance} & 
                \multicolumn{1}{c}{$\gamma_0$} & \multicolumn{1}{c}{$R_0$ (GV)} & \multicolumn{1}{l}{$s_0$} &
                \multicolumn{1}{c}{$\gamma_1$} & \multicolumn{1}{c}{$R_1$ (GV)} & \multicolumn{1}{l}{$s_1$} &  
                \multicolumn{1}{c}{$\gamma_2$} & \multicolumn{1}{c}{$R_2$ (GV)} & \multicolumn{1}{l}{$s_2$} &  
                \multicolumn{1}{c}{$\gamma_3$} & \multicolumn{1}{c}{$R_3$ (GV)} & \multicolumn{1}{l}{$s_3$} &  
                \multicolumn{1}{c}{$\gamma_4$}
                 }
	\startdata
$^{4}$He & standard & $1.026\times10^5$ & 2.10 & 1.0 & 0.20 & 1.81 & 7.5 & 0.30 & 2.41 & 340 & 0.13 & 2.12 & 30000 & 0.10 & 2.37\\
$^{3}$He (prim.) & standard & 295 & 0.50 & 8.5 & 0.29 & 1.0 & 12 & 0.20 & 2.41 & \nodata & \nodata & \nodata & \nodata & \nodata & \nodata \\
$^{4}$He & alternative & $1.60\times10^5$ & 2.05 & 1.0 & 0.26 & 1.85 & 7.5 & 0.27 & 2.51 & 350 & 0.17 & 2.14 & 30000 & 0.10 & 2.37\\
	\enddata
  \tablecomments{The $^{3}$He (prim.) row shows the injection spectrum for a primary $^{3}$He component, see Sect.~\ref{prim_3He}. The source abundances are relative, see Table 3 in \citet{2020ApJS..250...27B}. For parameter definitions see Eq.~(\ref{eq:1}). Shown are $|s_i|$ values, note that $s_i$ is negative/positive for $|\gamma_i |\lessgtr |\gamma_{i+1} |$.}
\end{deluxetable*}

Note that the breaks in the injection spectrum are phenomenological to provide the agreement with CR measurements at 1 AU. Here is our justification \citep{Moskalenko:2023xq}: While the supernova remnants (SNRs) are widely recognized as the main sources of CRs, the isotopic and spectral anomalies observed recently force us to look at other sources, especially local, which can also contribute to the observed CRs. These are primarily Wolf-Rayet stars (currently 354 are known) and O-stars (20,000 observed), which over their fairly short lifetimes provide, respectively, $10^{51}$ erg and $10^{50}$ erg in high-velocity winds reaching $(2-4)\times10^3$ km/s, pulsars ($\sim$1,500 observed) with their total rotating power reaching $4\times10^{49}$ erg (Crab), and novae providing $10^{45}$ erg (estimated frequency 30--50/year, $\approx$350 observed). For comparison, countless stellar flares can provide up to $10^{36}$ erg each, and can also add to low-energy CRs. Therefore, there is no reasonable way to derive the injection spectrum from the first principles, especially at low rigidities.  

\section{Results and discussion} \label{results}

The standard model results using the propagation parameters from Table~\ref{tbl-prop} are shown in Figs.~\ref{fig:He-spec-default} and \ref{fig:He-spec-default1} (left). The calculated spectra of total He and $^4$He reproduce the data quite well. Meanwhile, the difference between the PAMELA \citep{2016JPhCS.675c2001M} and AMS-02 \citep{2019PhRvL.123r1102A} data for $^4$He and $^3$He fluxes in the overlapping region \emph{increases} with rigidity, while the solar modulation effect \emph{decreases} with rigidity$-$the modulated spectra for appropriate modulation levels are shown by the solid red lines. This indicates possible systematic issues with the highest rigidity data points (2$-$3 GV) by PAMELA, which is a much smaller instrument than AMS-02\footnote{
The AMS-02 experiment uses modern technology, it has a stronger magnet, its response was thoroughly simulated and tested, and it also has several independent systems that allow for data cross-checks. Besides, the significantly larger acceptance area of AMS-02 and much longer exposure provide incomparable statistical accuracy. In turn, the large statistics combined with several detection systems, which can be used independently for cross checks, improve the rejection power of the instrument and its accuracy.
}.

At the same time, the predicted spectrum of $^3$He in the standard calculation (Table~\ref{tbl-inject}, see also \citealt{2020ApJS..250...27B}) exhibits a significant excess above 7~GV. This may be an indication for the presence of an additional source of these particles besides the fully secondary $^3$He produced in fragmentations of $Z\ge2$ nuclei or something else.

We see three possibilities, which may be responsible for the observed $^3$He excess: (i) the accuracy of the He production and fragmentation cross sections, (ii) the primary $^3$He component with a harder injection spectrum, and (iii) a non-uniform propagation probed on different spatial scales by light isotopes ($^{3,4}$He) and heavier species (e.g., the B/C ratio). Below we discuss these possibilities in detail.

\begin{figure*}[tbp!]
	\centering
	\includegraphics[width=0.5\textwidth]{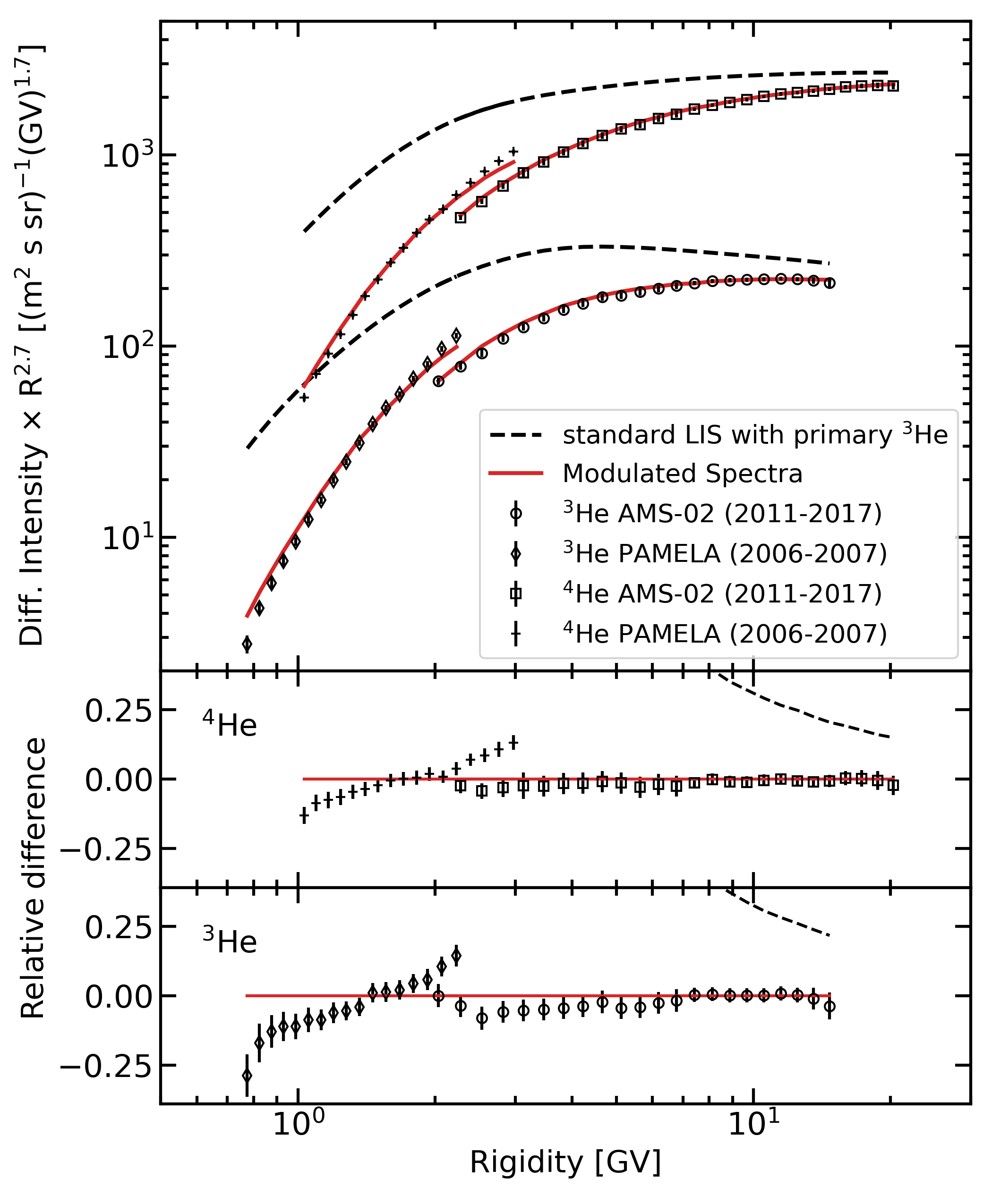}\\ 
	\includegraphics[width=0.5\textwidth]{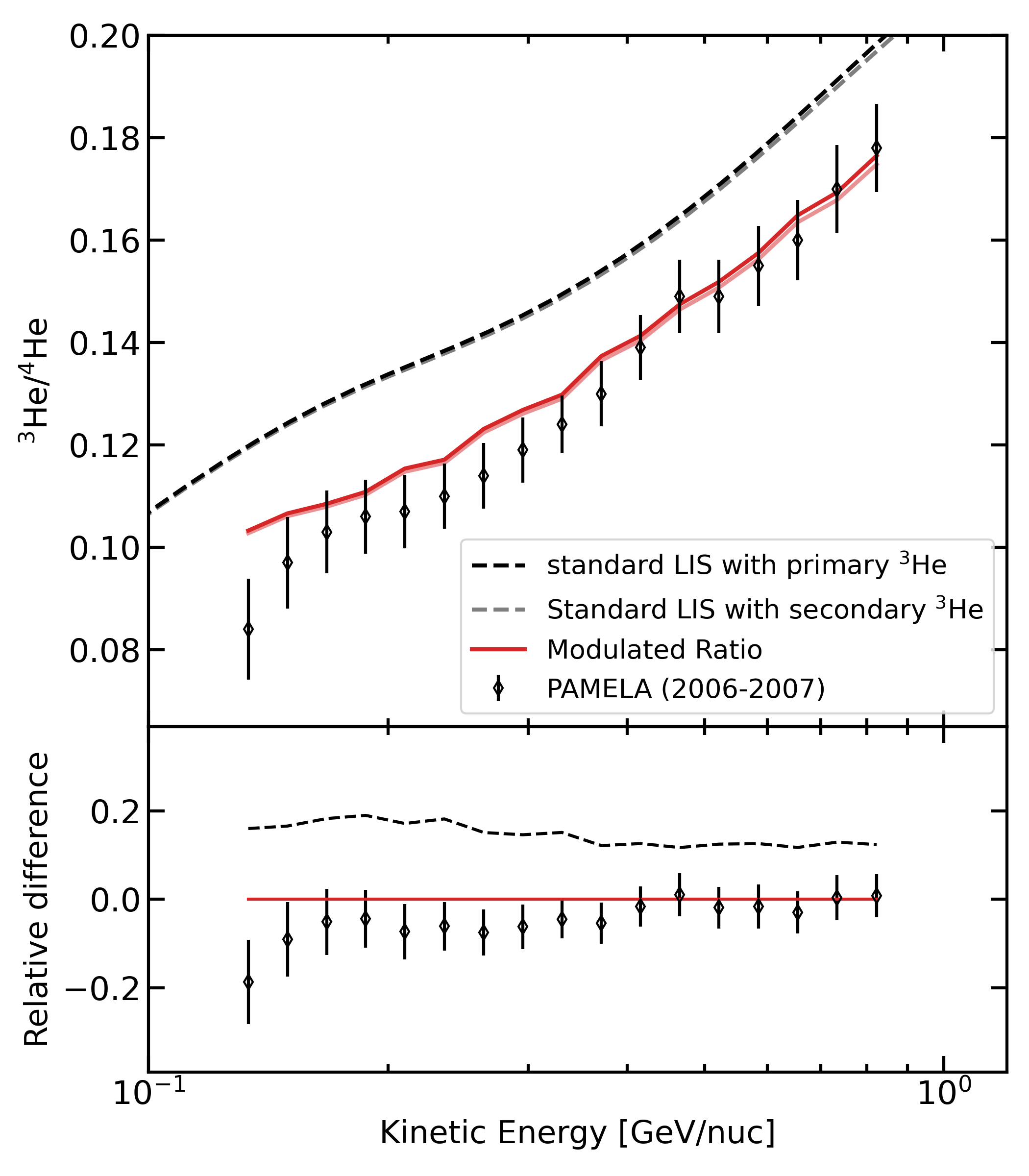}\hfill
	\includegraphics[width=0.5\textwidth]{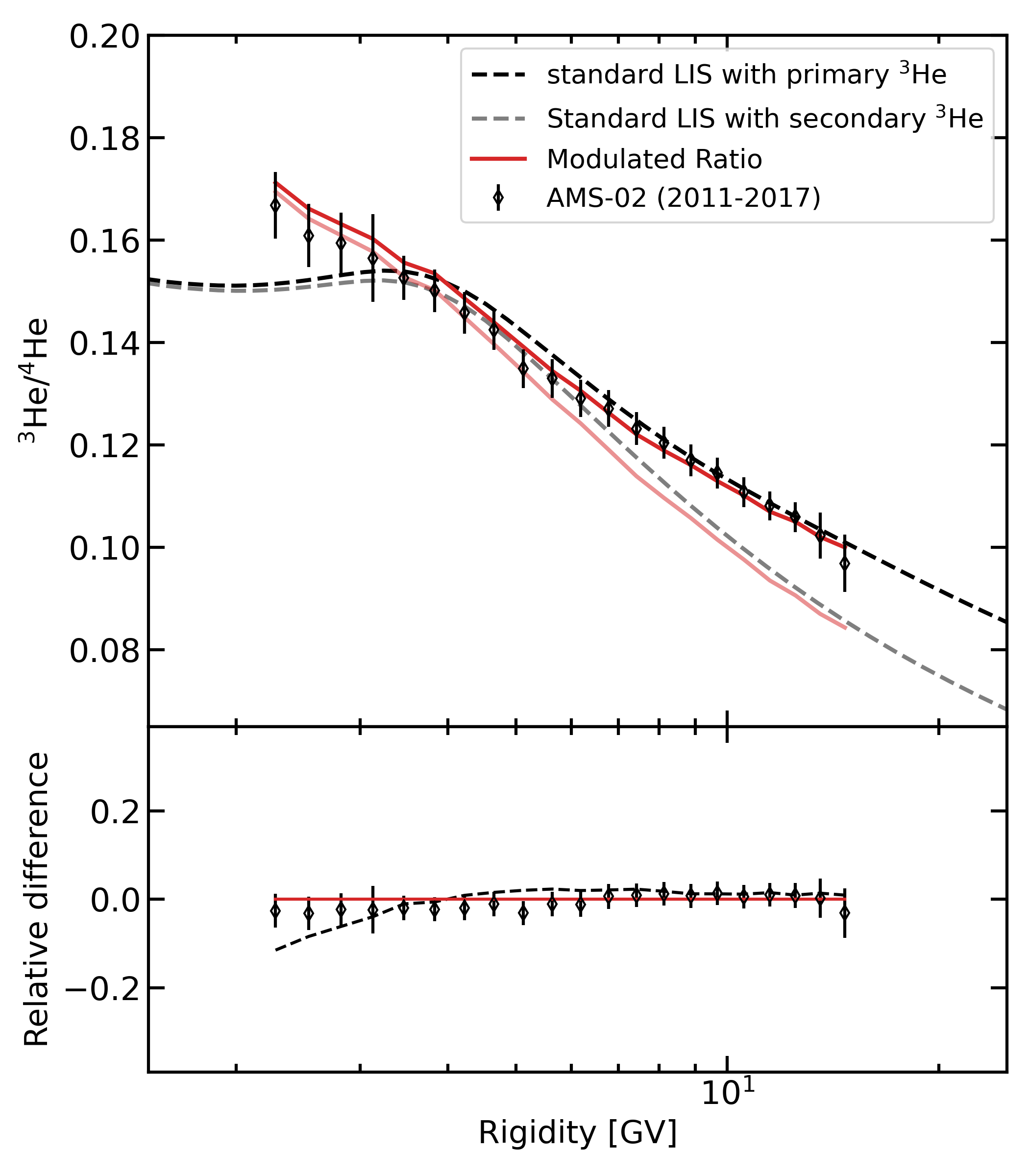} 
	\caption{
Calculations with the primary $^3$He component. \textit{Upper panel} shows a comparison of the calculated $^{3,4}$He spectra with with PAMELA \citep{2016JPhCS.675c2001M} and AMS-02 \citep{2019PhRvL.123r1102A} data. The dashed black lines show the LIS, and the solid red lines$-$the corresponding modulated spectra. \textit{Lower panels} show a comparison of the $^3$He/$^4$He ratio with PAMELA and AMS-02 data for two cases: fully secondary $^3$He as in Fig.~\ref{fig:He-spec-default} and with primary $^3$He component. Note different units in PAMELA and AMS-02 plots. The relative differences between our calculations and the data sets are shown in the bottom part of each panel. 
	}
	\label{fig:Prim3He}
\end{figure*}

\begin{figure*}[tbp!]
	\centering
	\includegraphics[width=0.5\textwidth]{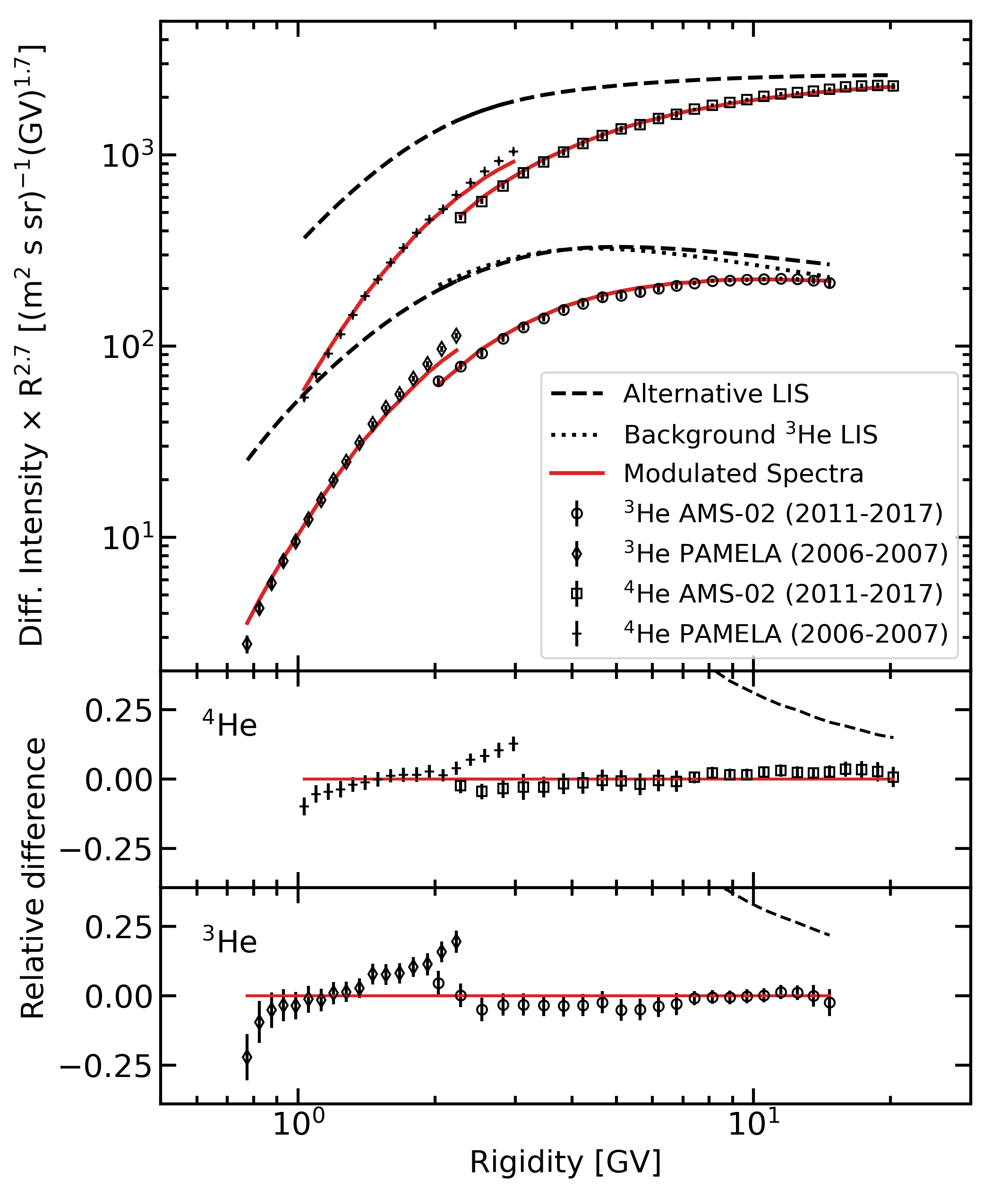}\\
	\includegraphics[width=0.5\textwidth]{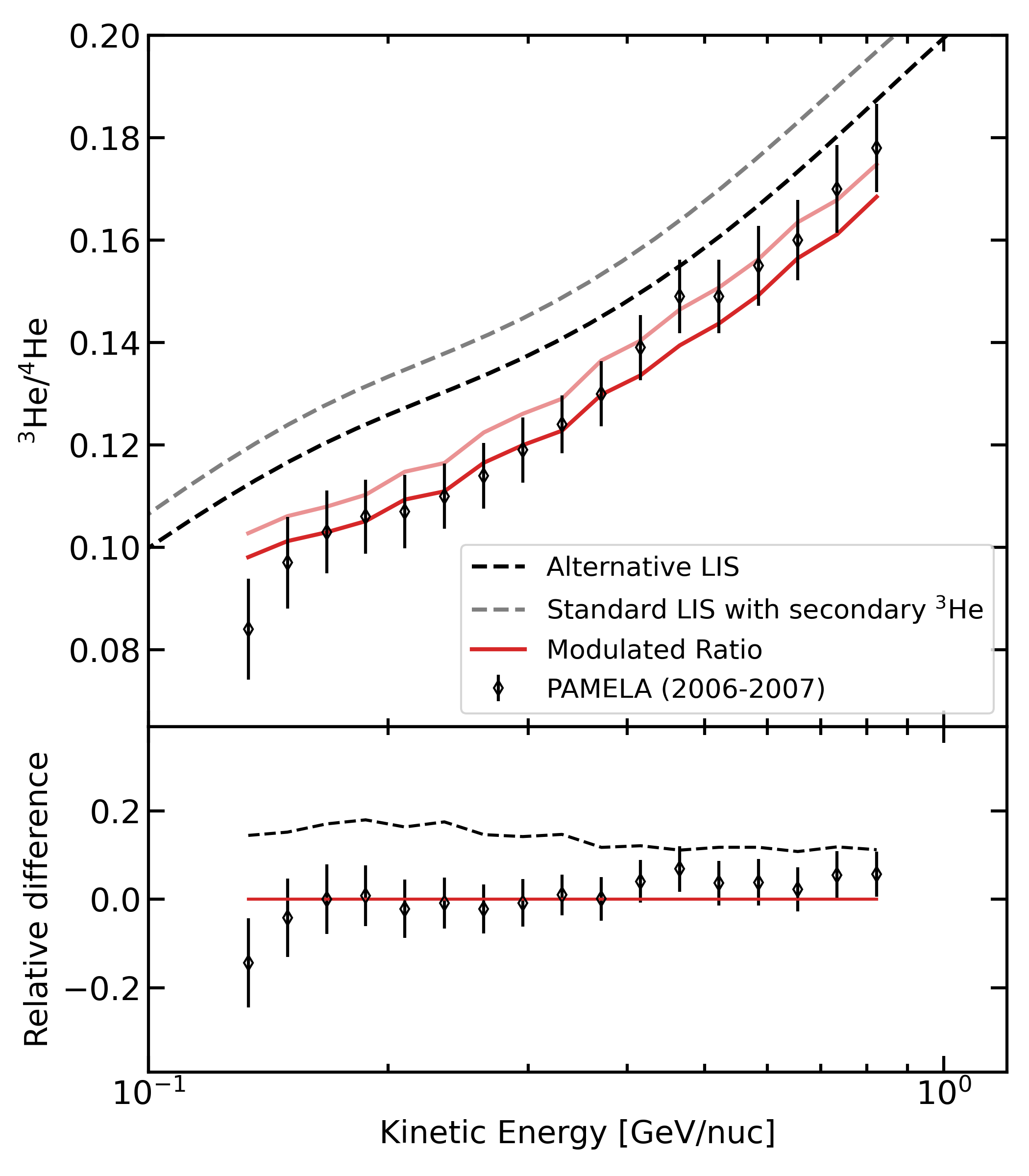}\hfill
	\includegraphics[width=0.5\textwidth]{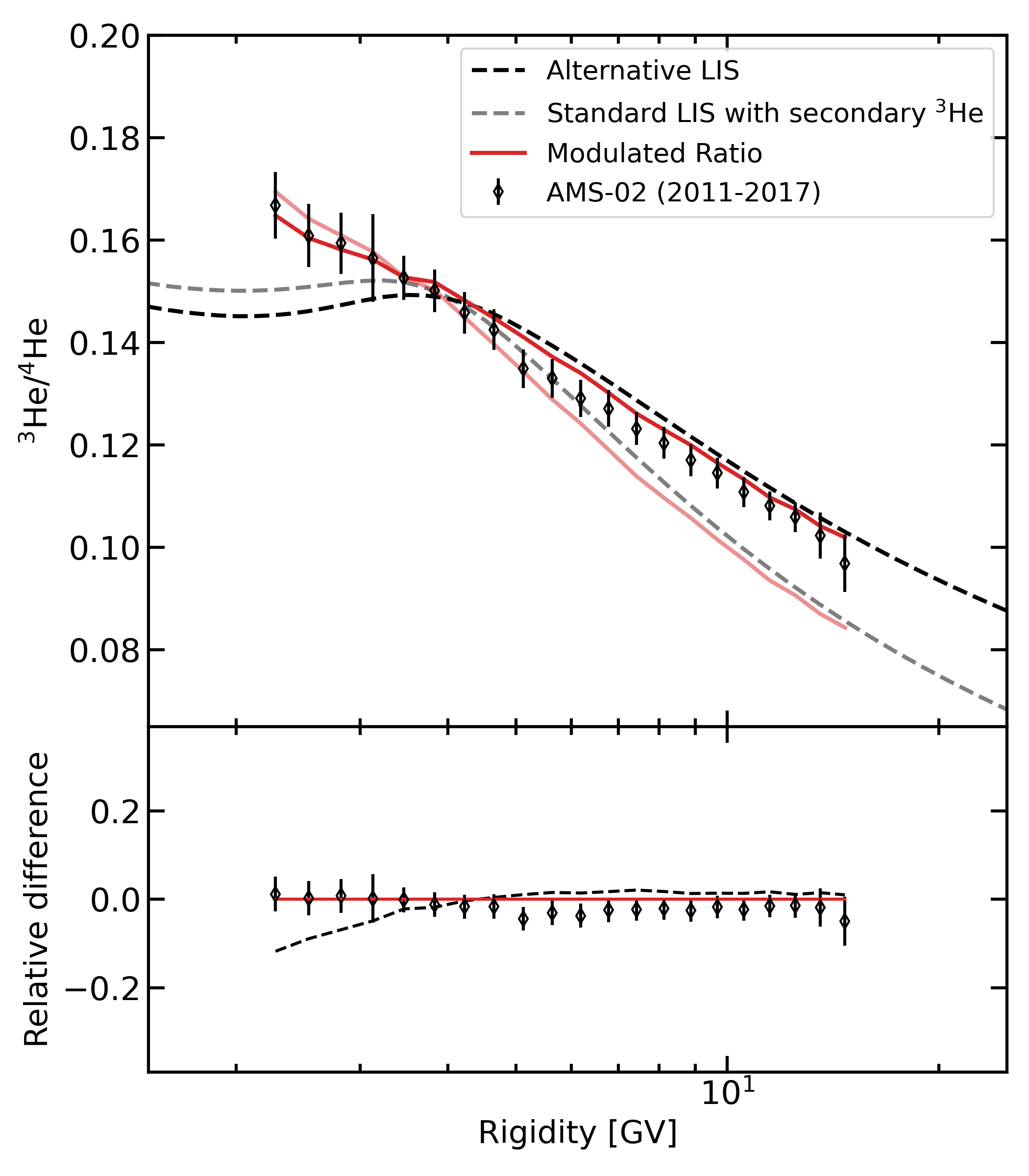} 
	\caption{
Calculations in the alternative model based on the measured $^3$He/$^4$He ratio. \textit{Upper panel} shows a comparison of the calculated $^{3,4}$He spectra with PAMELA \citep{2016JPhCS.675c2001M} and AMS-02 \citep{2019PhRvL.123r1102A} data. The dashed black lines show the LIS, and the solid red lines are the corresponding modulated spectra. The dotted line shows a contribution to the $^3$He from fragmentation of the $Z>2$ species calculated in the standard model (the so-called background $^3$He LIS). \textit{Lower panels} show a comparison of the calculated $^3$He/$^4$He ratio in the alternative and standard models with PAMELA and AMS-02 data. Note different units in PAMELA and AMS-02 plots. The relative differences between our calculations and the data sets are shown in the bottom part of each panel. 
	}
	\label{fig:Diffusion}
\end{figure*}

\subsection{The accuracy of cross sections hypothesis}

The accuracy of the $^3$He production and nuclear fragmentations cross sections were significantly improved in the latest version of \galprop{}. Moreover, the basic energy (rigidity) dependence of the relevant cross sections and of the discussed $^3$He excess provide an additional argument against this hypothesis.

A fundamental property of the production and fragmentation cross sections is their energy independence above $\sim$1$-$2 GeV/n. One can see the examples of the total fragmentation cross sections of different isotopes in \citet{2022ApJS..262...30P}, Figs.~10, 11. Examples of the $^3$He production cross sections in reactions with $^4$He and heavier species can be seen in Figs.~12, 15 of the same paper. Although some of the mentioned plots lack data points in the relevant energy range, the proposed approximations are consistent with a number of other cross section measurements.

For the flat cross sections, the slope of the $^3$He/$^4$He ratio depends only on the power-law index of the diffusion coefficient and the Alfv\`en speed. The latter makes the slope of the $^3$He/$^4$He ratio (2$-$20 GV) somewhat steeper than it is dictated by the rigidity dependence of the diffusion coefficient (Fig.~\ref{fig:He-spec-default}, lower right panel), while the effect of the solar modulation on the $^3$He/$^4$He ratio at $\gtrsim$3 GV is weak. 

The excess in the $^3$He spectrum and the $^3$He/$^4$He ratio is observed above 7 GV, which corresponds to $\gtrsim$3.8 GeV/n for $^3$He, and $\gtrsim$2.7 GeV/n for $^4$He and for other most abundant heavier projectiles ($A/Z$$\approx$2), such as $^{12}$C, $^{16}$O, $^{24}$Mg, $^{28}$Si. In the same figure, one can see a good agreement with standard calculations below 7 GV, which also includes the rigidities where the production and fragmentation cross sections have already become \emph{energy-independent (rigidity-independent)}, and the flattened $^3$He/$^4$He ratio above this rigidity, i.e.\ \emph{rigidity-dependent} excess. Therefore, the \emph{energy-independent (rigidity-independent)} cross sections cannot be the reason of the observed \emph{rigidity-dependent} $^3$He excess. The latter rules out the hypothesis of the accuracy of the cross sections.

\subsection{The hard spectrum component hypothesis}\label{prim_3He}

The second hypothesis implies that there is a hard spectrum component, which is either accelerated by a source enriched with $^3$He (primary $^3$He) relative to an ``average'' Galactic CR source, or is produced through $^4$He fragmentation in/near the source. Figs.~\ref{fig:He-spec-default1} (right) and \ref{fig:Prim3He} show calculations with the primary $^3$He, where we used the standard parameters listed in Table~\ref{tbl-prop}. For a comparison, the lower right panel in Fig.~\ref{fig:Prim3He} shows also the ratio for fully secondary $^3$He as in Fig.~~\ref{fig:He-spec-default}. The injection spectra of $^{3,4}$He isotopes are tuned to the AMS-02 data and provided in Table~\ref{tbl-inject}.

Although this case produces an excellent fit to the data, the idea of additional production of $^3$He through $^4$He fragmentation in/near the source has significant drawbacks. The most serious issue arises from the fact that the $^{3,4}$He fragmentation cross sections and the cross section for production of $^3$He in the reaction $^4$He$+p\to ^3$He are quite small, and much smaller than the cross sections for fragmentation of heavier species (C, N, O) and production of secondary isotopes of Be and B. Therefore, any process that effectively produces $^3$He by fragmentation of nuclei $Z\ge2$ produces Be and B even more efficiently given that C, N, O, and heavier species are also present in the CR source. 

However, such an overproduction of heavier species (Be, B) can be avoided in a scenario, where a local source is enriched with $^3$He relative to an ``average'' Galactic CR source. In this case, $^3$He is ``disconnected'' from heavier species and such scenario cannot be ruled out.

Indeed, the hypothesis of possible enrichment of a CR source environment with the $^3$He isotope looks plausible. The closest example is the solar energetic particle events. Some of them exhibit resonant enhancements of $^3$He/$^4$He up to 10,000-fold, which could even make $^3$He dominant over H in rare events. Even after fifty years of studies, the mechanism of this enhancement is not fully understood. A concise review of studies of $^3$He-rich events can be found in \citet{2021FrASS...8..164R}. Therefore, under certain circumstances, a similar mechanism could inject the $^3$He-rich material into the ISM, where it can be picked by a propagating shock, or inject it directly into the shock at the CR accelerator. While it is not clear how much such events could contribute to the particle spectra emergent from a source, it is evident that some level of enrichment of $^3$He by similar events may be plausible.

\subsection{The inhomogeneous diffusion hypothesis}

The third hypothesis is that the properties of the ISM are not uniform throughout the Galaxy, and that the CR propagation parameters vary across different spatial scales as a consequence. The latter can be tested using the secondary/primary nuclei ratios with significantly different inelastic cross sections \citep{2016ApJ...824...16J}, such as $\bar{p}/p$, $^3$He/$^4$He, B/C (or B/O), F/Ne (or F/Si), (Sc+V)/Fe. In the ratios $\bar{p}/p$ and $^3$He/$^4$He, the involved species have the smallest inelastic cross sections and thus can be used to test the propagation parameters over the largest Galactic volume, while (Sc+V)/Fe over the smallest. However, the low-energy feature (excess) found recently in the CR iron spectrum \citep{2021ApJ...913....5B} presumably associated with the local SNR event(s) 1--2 Mya, may hamper a straightforward interpretation of the (Sc+V)/Fe data.

Fig.~\ref{fig:Diffusion} shows calculations with propagation parameters derived from the $^3$He/$^4$He ratio itself assuming fully secondary $^3$He (Table~\ref{tbl-prop}, ``alternative model'')\footnote{
Currently, there is no way to derive the spatial dependence of propagation parameters from the first principles. Observations of CRs could provide such information. Meanwhile, the interpretation of the observed CR spectra may depend on the assumed propagation model.}. 
To do so, first, we performed a calculation of secondary $^3$He (the so-called ``background'') produced by fragmentation of nuclei $Z>2$ with fixed standard propagation parameters (Table~\ref{tbl-prop}, ``standard model'') and source abundance of $^4$He set to 0. Second, we performed an MCMC scan to derive the alternative propagation parameters with fixed halo size $z_h$$=$$4$ kpc, free source abundance of $^4$He, and abundances of nuclei $Z>2$ set to 0. The spectrum of $^3$He was calculated as a sum of the background, the fixed secondary component calculated in the first step, plus secondary $^3$He produced in fragmentation of $^4$He with alternative propagation parameters. The alternative propagation parameters derived in this calculation are shown in Table~\ref{tbl-prop} together with standard parameters for a side-by-side comparison.

The observation of the $^3$He/$^4$He excess over the calculation with standard propagation parameters, which is also observed in the $\bar{p}/p$ ratio \citep{2002ApJ...565..280M, 2003ApJ...586.1050M, 2006ApJ...642..902P}, implies an effectively smaller diffusion coefficient, at least below $\approx$20 GV, over the Galactic volume that is probed by it in a model-independent way---provided that we have similar scenarios in cases of lighter ($^3$He) and heavier (B) secondary species (CR source distribution, injection spectra, halo size $z_h$ etc.). Such an idea was first proposed by \citet{2016ApJ...824...16J}. Indeed, their analysis showed that the propagation parameters that best-fit $\bar{p}/p$ and He data are significantly different from those that fit heavier elements, including the B/C and $^{10}$Be/$^9$Be secondary-to-primary ratios normally used to calibrate the parameters of CR propagation models. The \galprop{} version 55 was used for that study along with data available at that time: ACE-CRIS, ISOMAX, PAMELA, HEAO-3, CREAM-I, II. AMS-02 data was not available at the time that the lengthy model scanning made by \citet{2016ApJ...824...16J} was begun.

More recent analyses using the $^3$He/$^4$He ratio and other AMS-02 data \citep{2019PhRvL.123r1102A, 2021PhR...894....1A} were performed by \citet{2022PhRvD.105j3033K} with the v56 \galprop{} release, although both versions 55 and 56 did not include the latest cross section updates, and by \citet{2023PhRvD.107l3008G} who also employed \galprop{} v56, but in that work the $^2$H and $^3$He production and fragmentation cross sections were modified according to \citet{2012A&A...539A..88C}.
The former analysis confirms the difference in the propagation parameters derived using different groups of nuclei, while the latter found an excess in $^3$He above 5$-$6 GeV/n over the calculations with the propagation parameters derived from the B/C ratio \citep{2020ApJS..250...27B}$-$in agreement with our conclusions.

\pagebreak
\subsection{The outcome}

Table~\ref{tbl-prop} shows that the propagation parameters for the standard and alternative models are different. To make a consistent comparison, the halo size in the alternative model was fixed to the same value $z_h=4$ kpc derived in the standard model. The diffusion coefficient normalization $D_0$ is larger for the alternative model, but both normalizations are consistent within the error bars. Significantly different are the indices of the diffusion coefficient $\delta$ and convection gradients $dV_{\rm conv}/dz$. The index of the diffusion coefficient in the alternative model $\delta=0.19$ is two times smaller than that in the standard model $\delta=0.415$, which implies a flatter $^3$He/$^4$He ratio, as observed. The convection gradient affects the spectral shape at low rigidities. The larger value for the standard model is responsible for a slight hardening of the $^3$He/$^4$He ratio below $\sim$4~GV. The alternative model matches the data at low rigidities and, therefore, an additional flattening is not required.  

In Appendix~\ref{Bartel} we provide a comparison of all three model predictions for the $^3$He and $^4$He fluxes with 21 AMS-02 data sets, each representing the averages over four Bartel rotations ($\approx$4 months). One can see that the standard model calculations show the excess in the $^3$He spectra above 7-8 GV for each plotted period, while $^4$He spectra agree well with the data (Figure~\ref{fig:Bartel-1}). The discrepancy is also reflected in the $\chi^2$/d.o.f.\ plot, which shows unacceptably large values for the $^3$He spectra. In the cases of a calculation with the primary $^3$He component (Figure~\ref{fig:Bartel-2}) and an alternative propagation model (Figure~\ref{fig:Bartel-3}) the agreement with data is significantly better and the $\chi^2$/d.o.f.\ for the $^3$He spectrum drops to values similar to $^4$He.

In Appendix~\ref{Tables}, we provide numerical tables for all three cases, which tabulate the isotopic $^{3,4}$He LIS and atomic He LIS in kinetic energy per nucleon $E_{\rm kin}$ and in rigidity $R$ in the standard model (Tables \ref{Tbl-He3LIS-EKin-default}$-$\ref{Tbl-HeliumLIS-Rigi-default}), $^{3}$He LIS and atomic He LIS for a model with the primary $^3$He component (Tables \ref{Tbl-He3LIS-EKin-prim}$-$\ref{Tbl-HeliumLIS-Rigi-prim}), and an alternative model with fully secondary $^3$He and adjusted propagation parameters (Tables \ref{Tbl-He3LIS-EKin-alternative}$-$\ref{Tbl-HeliumLIS-Rigi-alternative}). The tables for the alternative model are truncated near the spectral break rigidity at $\sim$300 GV. The $^{4}$He LIS is identical in the standard model and in a model with the primary $^3$He component, so it is provided only once in kinetic energy per nucleon $E_{\rm kin}$ and in rigidity $R$ variables.

\section{Discussion of other approaches}\label{alternatives} 

Some recent papers present claims for models that obtain consistency for the B/C and $^3$He/$^4$He ratios that use different approach. We elaborate on these further below. 

\citet{2020A&A...639A.131W} based their calculations on earlier papers \citep{2019PhRvD..99l3028G, 2019A&A...627A.158D, 2020A&A...639A..74W}.  The latter employ different options for parameterizing the isotopic production and total inelastic cross sections that are derived from those available with an earlier version of \galprop{}, transforming them using the so-called ``normalization, scale, and slope (NSS)'' method \citep{2019A&A...627A.158D}. This process modifies the normalization, energy scale, and shape of the cross sections through nuisance parameters, which are then tuned to the CR data using a sampling algorithm. At the same time, they also utilize the method of ``proxy reactions'', where a modified cross section of a single dominant channel for producing a given CR species is used to represent products over many channels. Additionally, the solar modulation is treated using the force-field approximation with modulation potential as another nuisance parameter. However, this method appears to need fine tuning for recovering even mock data \citep{2019A&A...627A.158D}. Still, the $^3$He/$^4$He ratio obtained by \citet{2020A&A...639A.131W} reproduces the data from $\approx$5$-$10 GeV/n but with a deficit from $\approx$1-2 GeV/n to 5 GeV/n (their Fig.\ 8). The cost of this approach is that there are uncontrolled modifications to the cross section parameterizations, which translate directly into the derived propagation parameters and calculated CR spectra.

\citet{2024PhRvD.109l3003D} use the latest version 57 of \galprop{}, but modify cross sections for production of secondary species ($^3$He, Li, Be, B) and secondary components in C and N. In particular, the normalizations $A_{\rm XS}$ and modifications of slopes $\delta_{\rm XS}$ of the production cross sections below 5 GeV/n are their nuisance parameters. The typical normalizations are $A_{\rm XS}$\,$^4$He$\to$$^3$He$\approx$1.2$-$1.3, $A_{\rm XS}$$\to$Li$\approx$1.3, $A_{\rm XS}$$\to$Be$\approx$1.05, $A_{\rm XS}$$\to$B$\approx$1.03$-$1.05, $A_{\rm XS}$$\to$C$\approx$0.4$-$1, and modifications of the slopes reach $\delta_{\rm XS}$\,$^4$He$\to$$^3$He$\approx$0.25, $\delta_{\rm XS}$$\to$Li$\approx$0.3, $\delta_{\rm XS}$$\to$Be$\approx$0.2, $\delta_{\rm XS}$$\to$B$\approx$0.13, $\delta_{\rm XS}$$\to$C$\approx$0.25 for some models. The largest adjustments are required for $^3$He and Li production cross sections, which support our findings of $^3$He and Li excesses \citep[for Li excess, see][]{2020ApJ...889..167B}. Interestingly, their calculated $^3$He/$^4$He ratio shows discontinuity at $\approx$5 GV (their Figs. 8, 9). 

These few papers stand out in the literature as employing the CR data as more reliable and use them to adjust cross section parameterizations, rather than the extensive nuclear physics data that are employed by most authors. While we do not agree with the methods employed by these works, that they need to significantly modify the cross section modeling to reconcile the different secondary to primary ratios in CRs ($^3$He, Li, Be, B) within a single propagation model supports our conclusions: i.e.\ that there is a difference in the propagation properties of the ISM on different scales, or that primary components in $^3$He and Li are needed.

\section{Conclusion} \label{conclusion}

Using the combined data of AMS-02 \citep{2021PhRvL.126d1104A}, PAMELA \citep{2016JPhCS.675c2001M}, ACE-CRIS \citep{2013ApJ...770..117L}, and Voyager 1 \citep{2016ApJ...831...18C} we analyzed the spectra of He and its isotopes $^{3,4}$He over a wide rigidity range $\ge$100 MV. We found that the He and $^4$He spectra agree well with the predictions made with the \galprop{}-\helmod{} framework, while $^3$He spectrum shows a significant excess above 7 GV in rigidity and up to 13 GV where the AMS-02 data available. 

We further show that this excess can be plausibly explained assuming a primary $^3$He component with a harder injection spectrum, or a non-uniform propagation probed on different spatial scales by light isotopes ($^{3,4}$He) and heavier species (e.g., the B/C ratio). Meanwhile, the accuracy of the He isotopic production and fragmentation cross sections as the main reason for the observed excess can be ruled out based on the energy (rigidity) dependences of the relevant cross sections and of the discussed $^3$He excess.

Finally, we note that the exploration of the newly discovered features in the spectra of CR species has just begun, thanks to the data from the interstellar probes Voyager 1, 2, and precise measurements by AMS-02, ACE-CRIS, and other instruments. These features harbor the keys to understanding our local Galactic environment and the history of formation of the solar system. The increase in the collected statistics and reduction in the systematic errors will help to establish the precise spectral shapes of observed features and to facilitate their interpretation. 

\acknowledgements
Special thanks to Pavol Bobik, Giuliano Boella, Marian Putis, and Mario Zannoni for their continuous support of the \helmod{} project and many useful suggestions. 
This work was carried out using HELMOD tool, which is currently supported within ASIF$-$ASI (Agenzia Spaziale Italiana) Supported Irradiation Facilities$-$framework for space radiation environment activities, e.g., ASIF implementation agreements 2017-15-HD.0 ASI-INFN,  2017-22-HD.0 ASI-ENEA, 2021-36-HH.0 ASI-Milano-Bicocca University, 2021-39-HH.0 ASI-ENEA and 2024-21-HH ASI-INFN. This work is also supported by ASI contract ASI-INFN 2019-19-HH.0 and ESA (European Space Agency) contract 4000116146/16/NL/HK.
Igor V.\ Moskalenko and Troy A.\ Porter acknowledge support from NASA Grants No.~80NSSC23K0169, 80NSSC22K0718, 80NSSC22K0477. This research has made use of the SSDC CR database \citep{2017ICRC...35.1073D} and LPSC Database of Charged CRs \citep{2014A&A...569A..32M}.

\bibliography{bibliography}

\appendix

\section{$^3$He and $^4$He fluxes averaged over Bartel rotations}\label{Bartel}

The reliability of our results can be verified through a comparison of all three model predictions for the $^3$He and $^4$He fluxes and the $^3$He/$^4$He ratios with the AMS-02 data averaged over four Bartel rotations ($\sim$4 months). In particular, we compare our results with Bartel rotation average fluxes for Bartel rotations: 2426$-$2429, 2430$-$2433, 2434$-$2437, 2438$-$2441, 2442$-$2445, 2446$-$2449, 2450$-$2453, 2454$-$2457, 2458$-$2461, 2462$-$2465, 2466$-$2469, 2474$-$2477, 2478$-$2481, 2482$-$2485, 2486$-$2489, 2490$-$2493, 2494$-$2497, 2498$-$2501, 2502$-$2505, 2506$-$2509, 2510$-$2513, as reported in \citet{2019PhRvL.123r1102A}. We also report the $\chi^2$/d.o.f.\ calculated for each Bartel rotation period for both $^3$He and $^4$He spectra.

\begin{figure}[tbh!]
	\centering
	\includegraphics[height=0.4075\textheight]{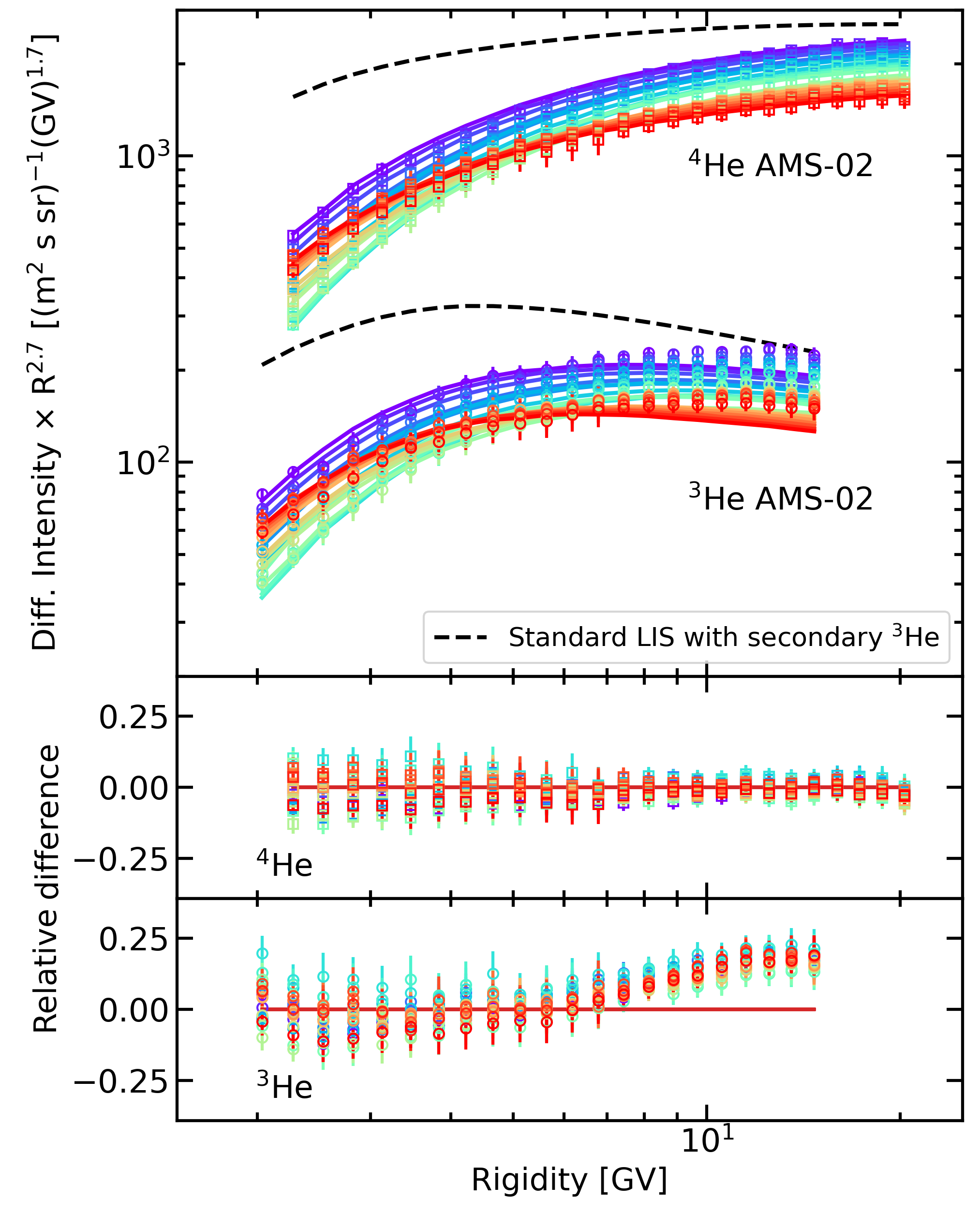}\hfill
	\includegraphics[height=0.4075\textheight]{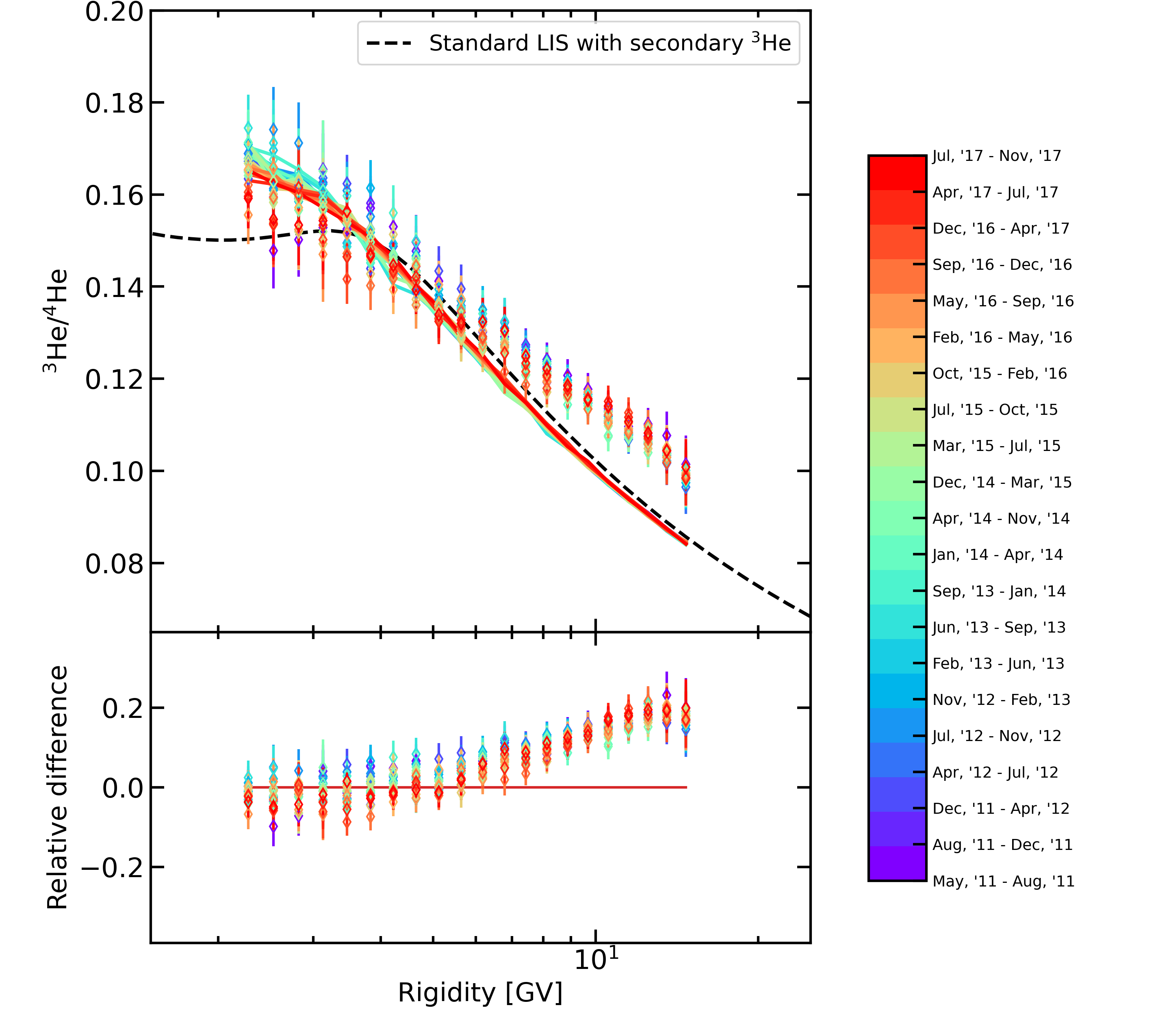}\\
	\includegraphics[width=0.7\textwidth]{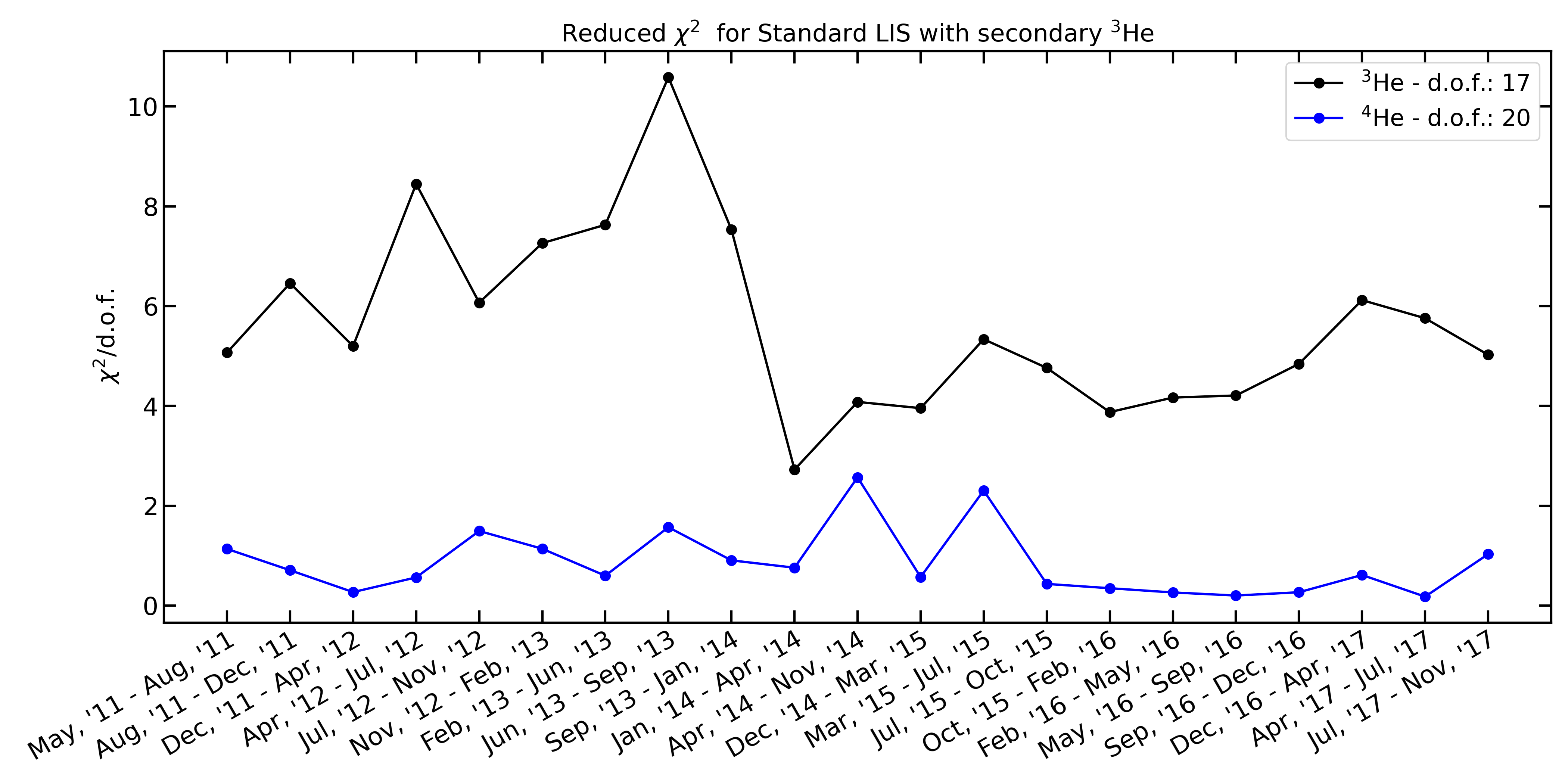}
	\caption{A comparison of the standard calculations for the $^3$He and $^4$He fluxes \emph{(upper left panel)} and the $^3$He/$^4$He ratio \emph{(upper right panel)} with the AMS-02 data \citep{2019PhRvL.123r1102A} averaged over four Bartel rotations. The dashed black lines show the LIS \emph{(left)} or the LIS ratio \emph{(right)}, and the solid colored lines$-$the corresponding modulated values. Note that the fluxes in the \emph{left panel} shown with colored lines and corresponding data points are renormalized with a factor $0.98^n$ each for clarity, here $n=0..20$ is the data set number. The color coding scales and the corresponding periods are shown in the right. The relative differences between our calculations and the data sets are shown in the bottom part of each panel. \emph{Lower panel} shows the $\chi^2$/d.o.f.\ calculated for each Bartel rotation period for both $^3$He and $^4$He spectra. 
	}
	\label{fig:Bartel-1}
\end{figure}

\begin{figure*}[tbh!]
	\centering
	\includegraphics[height=0.4075\textheight]{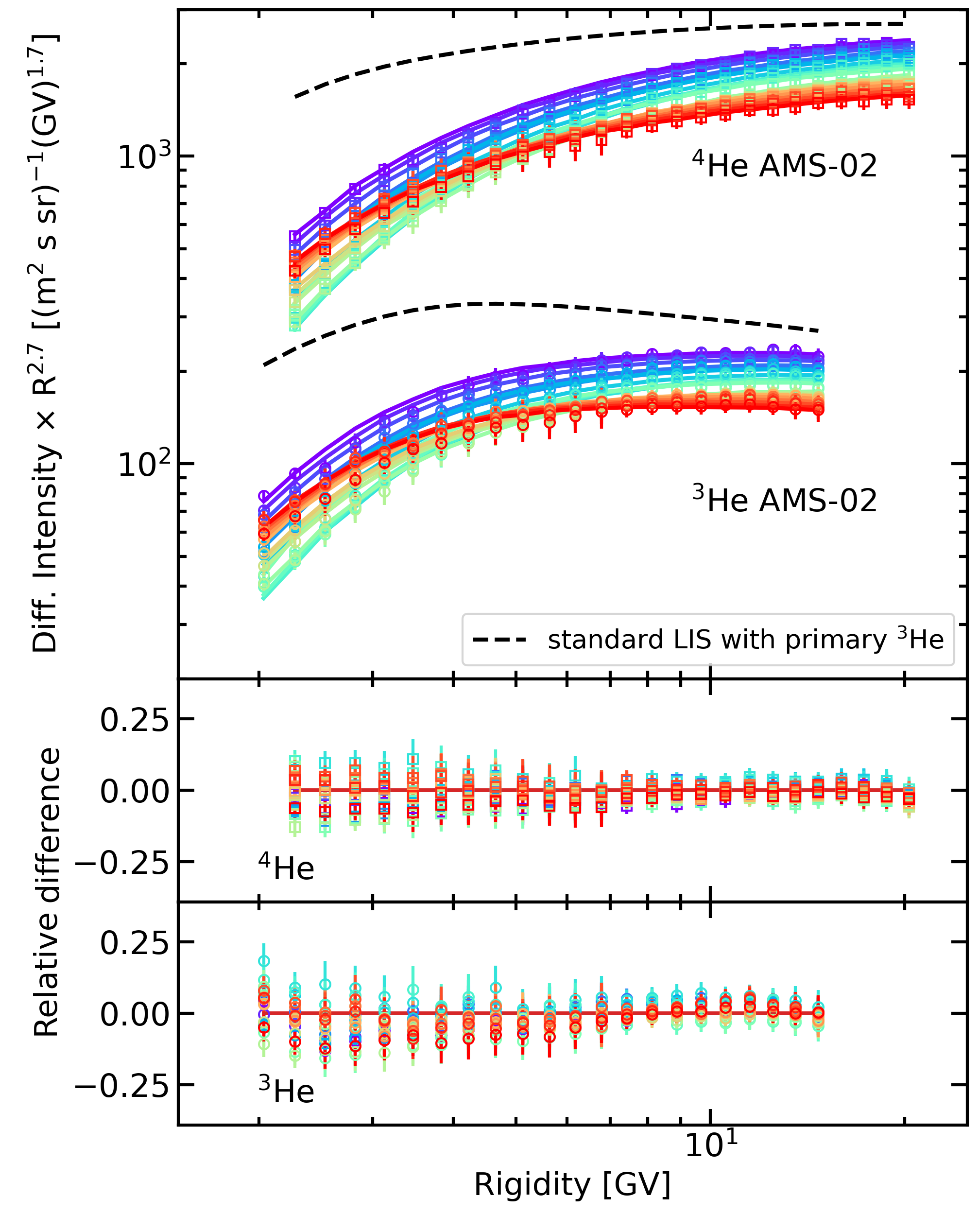}\hfill
	\includegraphics[height=0.4075\textheight]{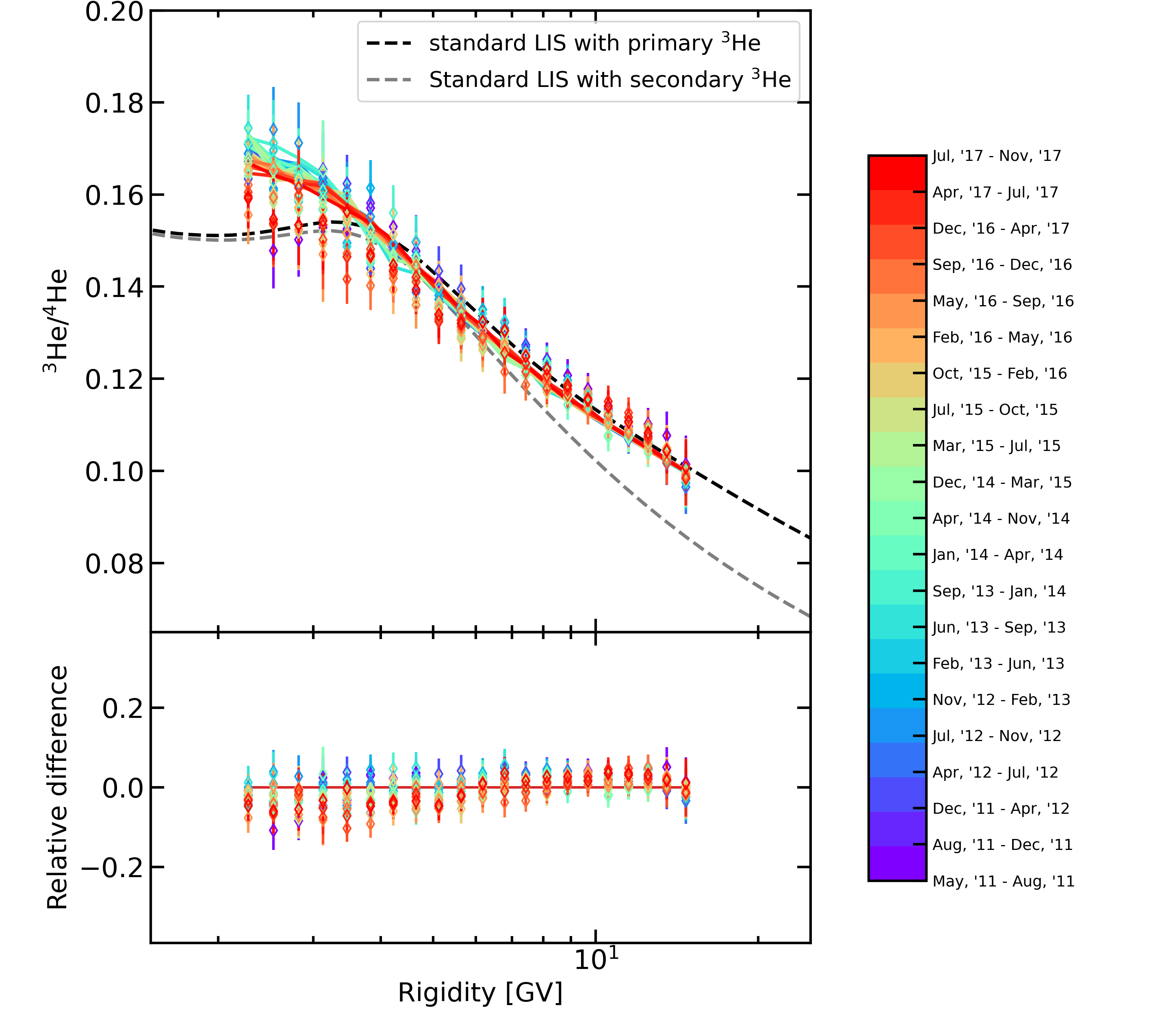}\\
	\includegraphics[width=0.7\textwidth]{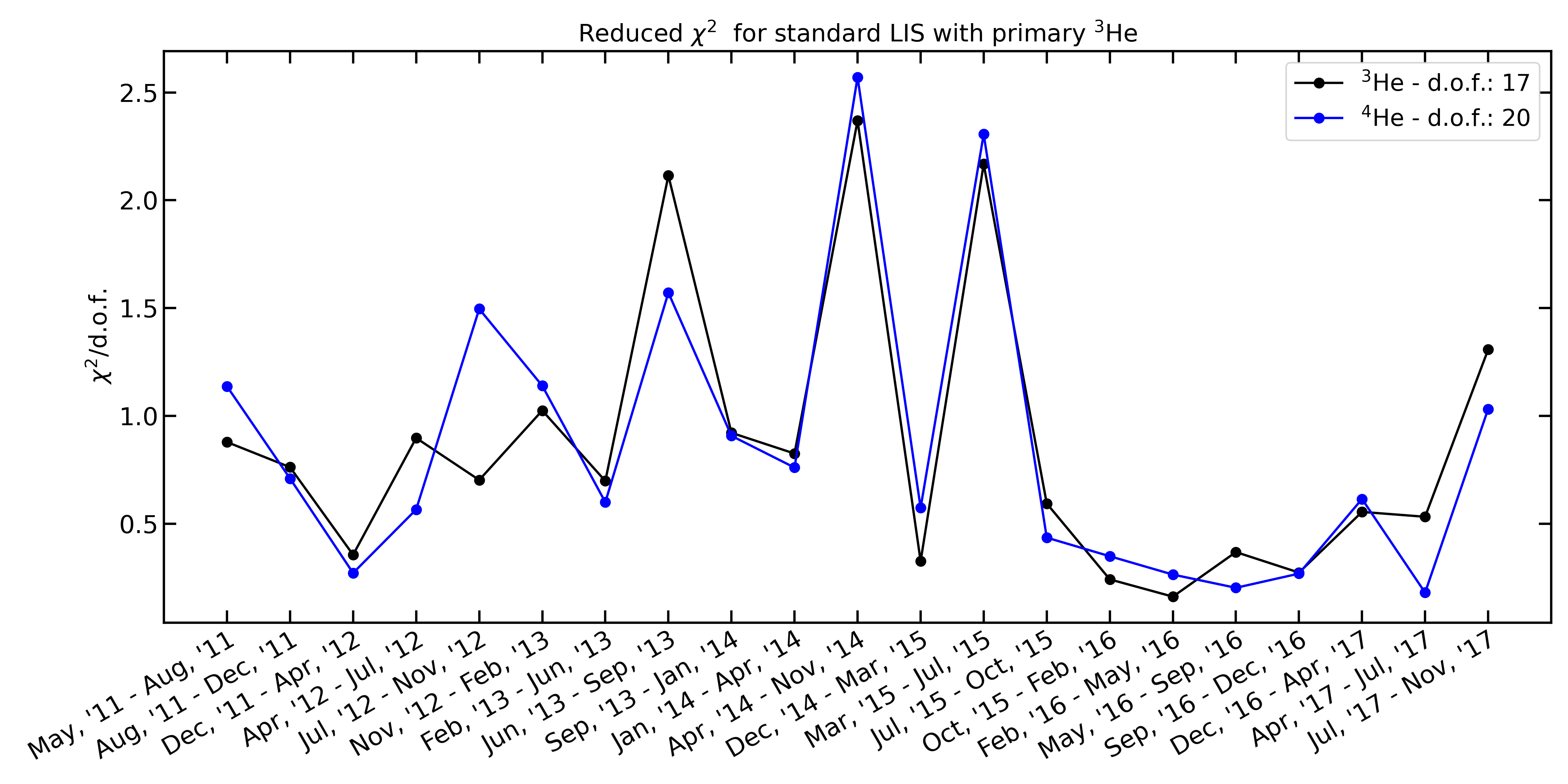}
	\caption{Calculations with the primary $^3$He component. Shown is a comparison of calculations for the $^3$He and $^4$He fluxes \emph{(upper left panel)} and the $^3$He/$^4$He ratio \emph{(upper right panel)} with the AMS-02 data \citep{2019PhRvL.123r1102A} averaged over four Bartel rotations. The dashed black lines show the LIS \emph{(left)} or the LIS ratio \emph{(right)}, and the solid colored lines$-$the corresponding modulated values. Note that the fluxes in the \emph{left panel} shown with colored lines and corresponding data points are renormalized with a factor $0.98^n$ each for clarity, here $n=0..20$ is the data set number. The color coding scales and the corresponding periods are shown in the right. The relative differences between our calculations and the data sets are shown in the bottom part of each panel. \emph{Lower panel} shows the $\chi^2$/d.o.f.\ calculated for each Bartel rotation period for both $^3$He and $^4$He spectra.  
	}
	\label{fig:Bartel-2}
\end{figure*}

\begin{figure*}[tbh!]
	\centering
	\includegraphics[height=0.408\textheight]{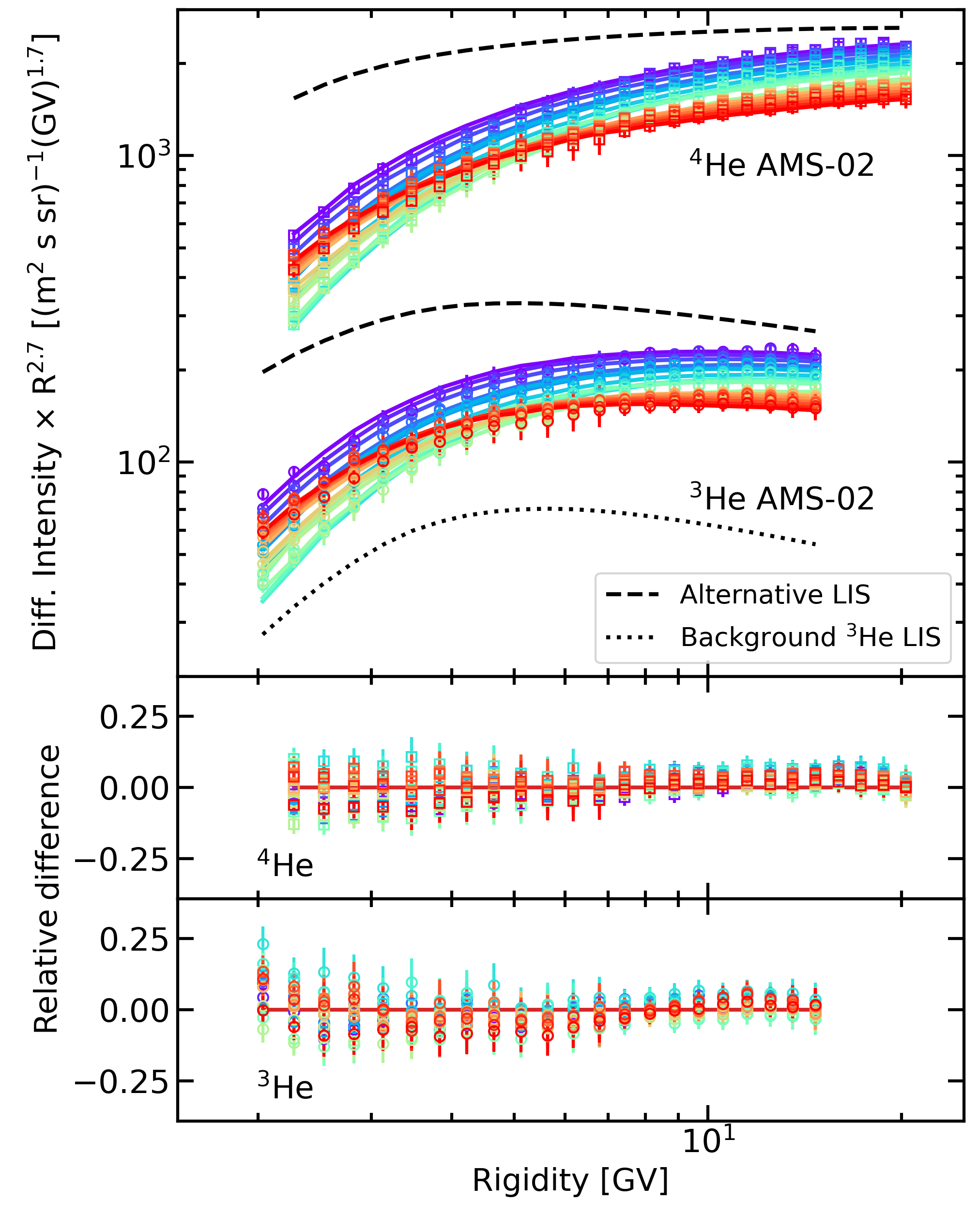}\hfill
	\includegraphics[height=0.408\textheight]{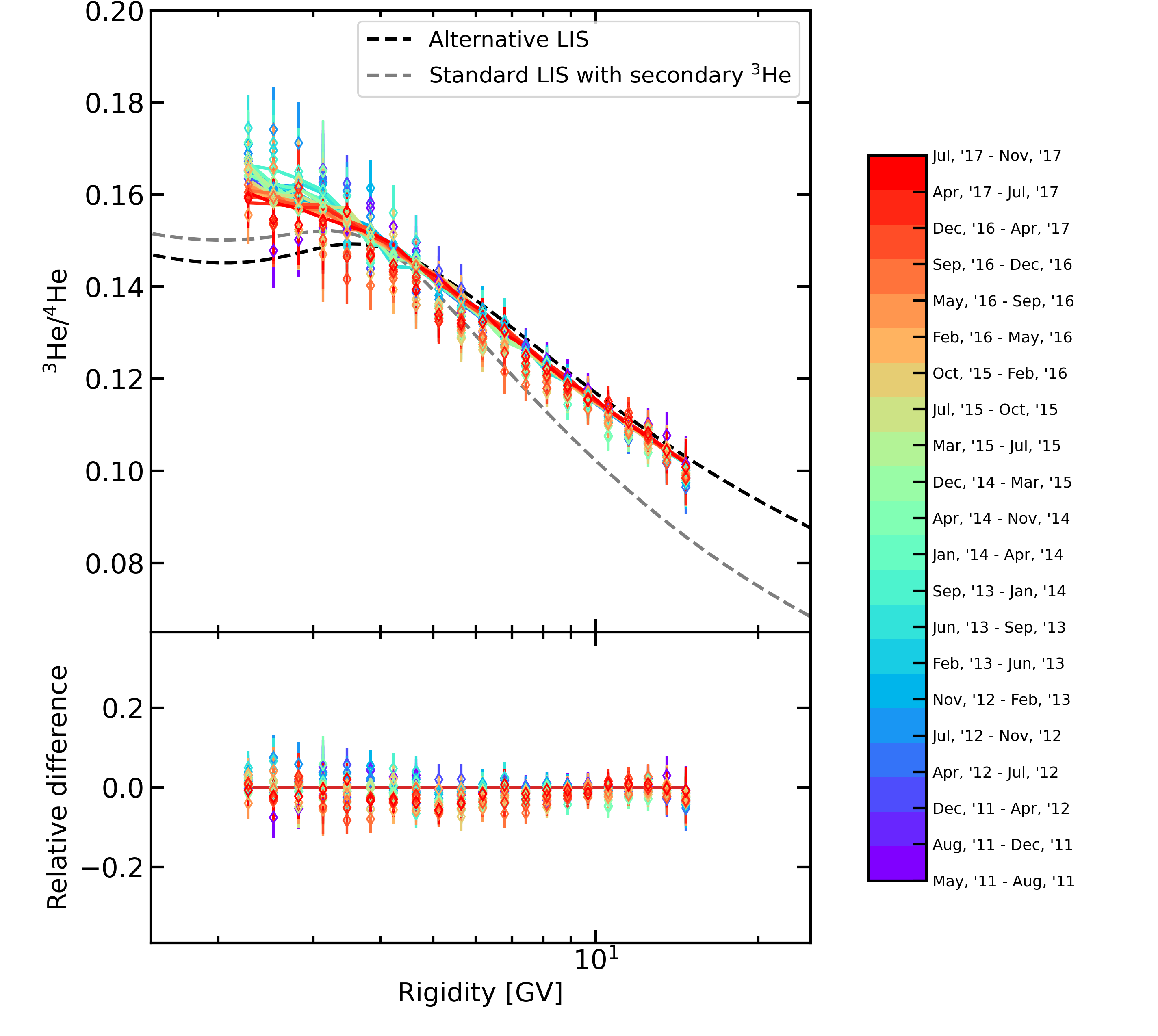}\\
	\includegraphics[width=0.7\textwidth]{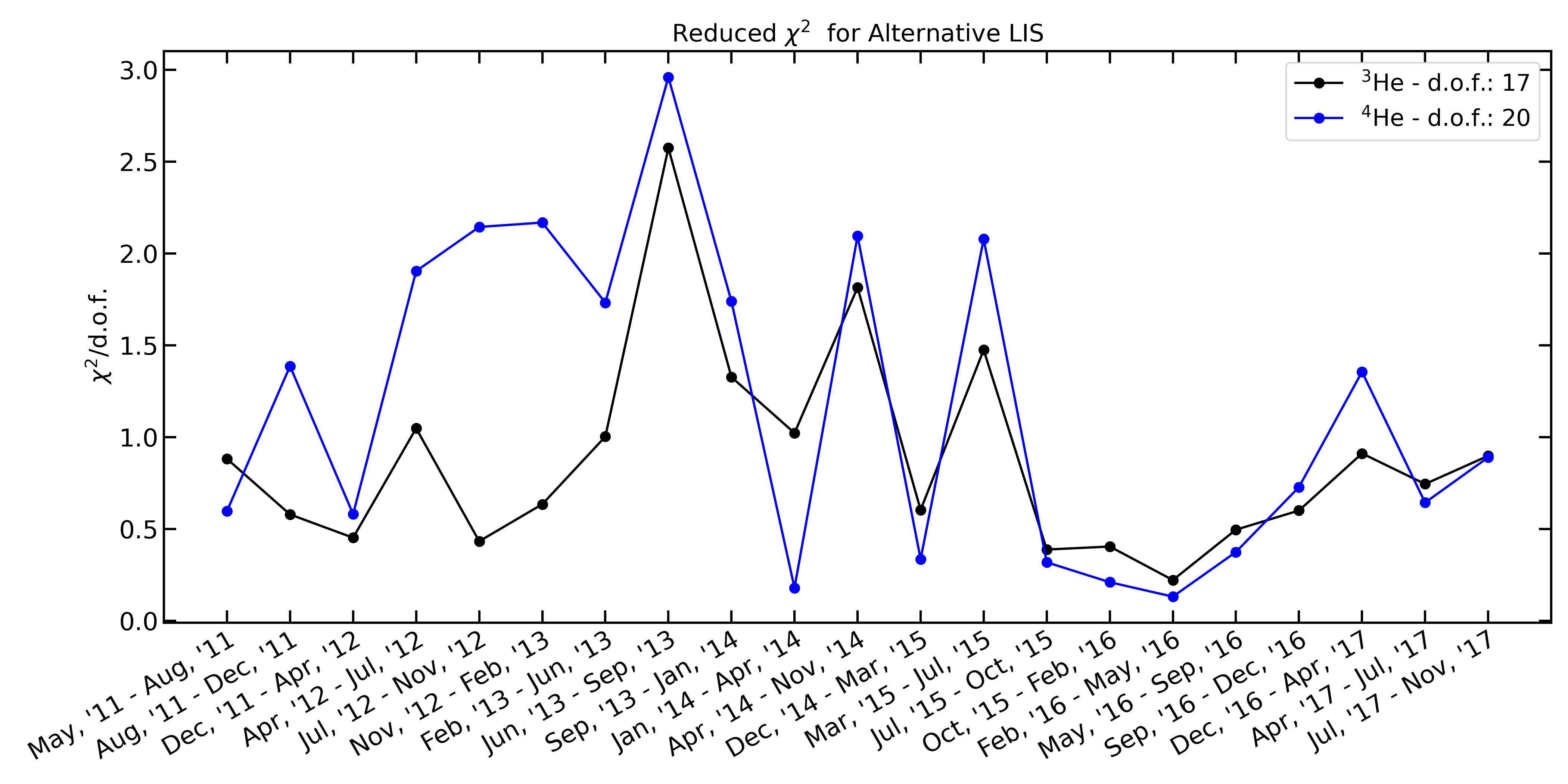}
	\caption{Calculations in the alternative model based on the $^3$He/$^4$He ratio. Shown is a comparison of calculations for the $^3$He and $^4$He fluxes \emph{(upper left panel)} and the $^3$He/$^4$He ratio \emph{(upper right panel)} with the AMS-02 data \citep{2019PhRvL.123r1102A} averaged over four Bartel rotations. The dashed black lines show the LIS \emph{(left)} or the LIS ratio \emph{(right)}, and the solid colored lines$-$the corresponding modulated values. Note that the fluxes in the \emph{left panel} shown with colored lines and corresponding data points are renormalized with a factor $0.98^n$ each for clarity, here $n=0..20$ is the data set number. The color coding scales and the corresponding periods are shown in the right. The relative differences between our calculations and the data sets are shown in the bottom part of each panel. \emph{Lower panel} shows the $\chi^2$/d.o.f.\ calculated for each Bartel rotation period for both $^3$He and $^4$He spectra. 
	}
	\label{fig:Bartel-3}
\end{figure*}

One can see that the standard model calculations show the excess in the $^3$He spectra above 7-8 GV for each plotted period, while $^4$He spectra agree well with the data (Figure~\ref{fig:Bartel-1}). The discrepancy is also reflected in the $\chi^2$/d.o.f.\ plot, which shows unacceptably large values for the $^3$He spectra. In the cases of a calculation with the primary $^3$He component (Figure~\ref{fig:Bartel-2}) and an alternative propagation model (Figure~\ref{fig:Bartel-3}) the agreement with data is significantly better and the $\chi^2$/d.o.f.\ for the $^3$He spectrum drops to values similar to $^4$He.

\section{Numerical tables of the $^{3,4}$He and atomic He LIS}\label{Tables}

Here we tabulate the LIS in kinetic energy per nucleon $E_{\rm kin}$ and in rigidity $R$, in the standard model, Tables \ref{Tbl-He3LIS-EKin-default}$-$\ref{Tbl-HeliumLIS-Rigi-default}, corresponding Tables \ref{Tbl-He3LIS-EKin-prim}$-$\ref{Tbl-HeliumLIS-Rigi-prim} for a model with the primary $^3$He component, and an alternative model with fully secondary $^3$He and adjusted propagation parameters, Tables \ref{Tbl-He3LIS-EKin-alternative}$-$\ref{Tbl-HeliumLIS-Rigi-alternative}. The tables for the alternative model are  truncated near the spectral break rigidity at $\sim$300 GV. The $^{4}$He LIS is identical in the standard model and in a model with the primary $^3$He component, so it is provided only once in kinetic energy per nucleon $E_{\rm kin}$ and in rigidity $R$ variables.


\begin{deluxetable*}{cccccccccc}
\tabletypesize{\footnotesize}
    \tablecolumns{10}
    \tablewidth{0mm}
    \tablecaption{Secondary $^3$He LIS, standard model \label{Tbl-He3LIS-EKin-default}}
    \tablehead{
        \colhead{$E_{\rm kin}$} & \colhead{Differential} &
        \colhead{$E_{\rm kin}$} & \colhead{Differential} &
        \colhead{$E_{\rm kin}$} & \colhead{Differential} &
        \colhead{$E_{\rm kin}$} & \colhead{Differential} &
        \colhead{$E_{\rm kin}$} & \colhead{Differential} \\ [-2ex] 
        \colhead{GeV/n} & \colhead{intensity} &
        \colhead{GeV/n} & \colhead{intensity} &
        \colhead{GeV/n} & \colhead{intensity} &
        \colhead{GeV/n} & \colhead{intensity} &
        \colhead{GeV/n} & \colhead{intensity}  
    }
\startdata
1.000e-03 & 5.691e+00 & 4.725e-02 & 1.526e+02 & 2.233e+00 & 8.580e+00 & 1.055e+02 & 1.402e-04 & 4.984e+03 & 1.731e-09\\
1.101e-03 & 6.520e+00 & 5.203e-02 & 1.596e+02 & 2.458e+00 & 6.929e+00 & 1.162e+02 & 1.044e-04 & 5.489e+03 & 1.306e-09\\
1.213e-03 & 7.073e+00 & 5.730e-02 & 1.659e+02 & 2.707e+00 & 5.556e+00 & 1.279e+02 & 7.787e-05 & 6.044e+03 & 9.851e-10\\
1.335e-03 & 7.661e+00 & 6.309e-02 & 1.715e+02 & 2.981e+00 & 4.430e+00 & 1.409e+02 & 5.812e-05 & 6.656e+03 & 7.426e-10\\
1.470e-03 & 8.298e+00 & 6.948e-02 & 1.761e+02 & 3.283e+00 & 3.511e+00 & 1.551e+02 & 4.343e-05 & 7.329e+03 & 5.595e-10\\
1.619e-03 & 8.992e+00 & 7.651e-02 & 1.798e+02 & 3.615e+00 & 2.767e+00 & 1.708e+02 & 3.250e-05 & 8.071e+03 & 4.213e-10\\
1.783e-03 & 9.745e+00 & 8.425e-02 & 1.824e+02 & 3.981e+00 & 2.169e+00 & 1.881e+02 & 2.435e-05 & 8.887e+03 & 3.170e-10\\
1.963e-03 & 1.056e+01 & 9.277e-02 & 1.840e+02 & 4.384e+00 & 1.693e+00 & 2.071e+02 & 1.826e-05 & 9.786e+03 & 2.384e-10\\
\enddata
    \tablecomments{Differential Intensity units: (m$^2$ s sr GeV/n)$^{-1}$.}
\end{deluxetable*}

\begin{deluxetable*}{cccccccccc}
\tabletypesize{\footnotesize}
    \tablecolumns{10}
    \tablewidth{0mm}
    \tablecaption{Secondary $^3$He LIS, standard model \label{Tbl-He3LIS-Rigi-default}}
    \tablehead{
        \colhead{Rigidity} & \colhead{Differential} &
        \colhead{Rigidity} & \colhead{Differential} &
        \colhead{Rigidity} & \colhead{Differential} &
        \colhead{Rigidity} & \colhead{Differential} &
        \colhead{Rigidity} & \colhead{Differential} \\ [-2ex] 
        \colhead{GV} & \colhead{intensity} &
        \colhead{GV} & \colhead{intensity} &
        \colhead{GV} & \colhead{intensity} &
        \colhead{GV} & \colhead{intensity} &
        \colhead{GV} & \colhead{intensity}  
    }
\startdata
6.476e-02 & 1.757e-01 & 4.506e-01 & 3.122e+01 & 4.536e+00 & 5.467e+00 & 1.596e+02 & 9.347e-05 & 7.478e+03 & 1.154e-09\\
6.796e-02 & 2.112e-01 & 4.735e-01 & 3.415e+01 & 4.889e+00 & 4.442e+00 & 1.756e+02 & 6.962e-05 & 8.234e+03 & 8.708e-10\\
7.132e-02 & 2.404e-01 & 4.975e-01 & 3.711e+01 & 5.276e+00 & 3.581e+00 & 1.933e+02 & 5.191e-05 & 9.067e+03 & 6.568e-10\\
7.484e-02 & 2.732e-01 & 5.229e-01 & 4.006e+01 & 5.700e+00 & 2.868e+00 & 2.127e+02 & 3.875e-05 & 9.985e+03 & 4.951e-10\\
7.854e-02 & 3.105e-01 & 5.496e-01 & 4.298e+01 & 6.165e+00 & 2.283e+00 & 2.341e+02 & 2.895e-05 & 1.099e+04 & 3.730e-10\\
8.242e-02 & 3.530e-01 & 5.778e-01 & 4.580e+01 & 6.675e+00 & 1.805e+00 & 2.576e+02 & 2.166e-05 & 1.211e+04 & 2.809e-10\\
8.649e-02 & 4.014e-01 & 6.075e-01 & 4.850e+01 & 7.235e+00 & 1.420e+00 & 2.835e+02 & 1.623e-05 & 1.333e+04 & 2.114e-10\\
9.077e-02 & 4.566e-01 & 6.389e-01 & 5.101e+01 & 7.849e+00 & 1.111e+00 & 3.121e+02 & 1.218e-05 & 1.468e+04 & 1.589e-10\\
\enddata
    \tablecomments{Differential Intensity units: (m$^2$ s sr GV)$^{-1}$.}
\end{deluxetable*}

\begin{deluxetable*}{cccccccccc}
\tabletypesize{\footnotesize}
    \tablecolumns{10}
    \tablewidth{0mm}
    \tablecaption{$^4$He LIS, standard model \label{Tbl-He4LIS-EKin-default}}
    \tablehead{
        \colhead{$E_{\rm kin}$} & \colhead{Differential} &
        \colhead{$E_{\rm kin}$} & \colhead{Differential} &
        \colhead{$E_{\rm kin}$} & \colhead{Differential} &
        \colhead{$E_{\rm kin}$} & \colhead{Differential} &
        \colhead{$E_{\rm kin}$} & \colhead{Differential} \\ [-2ex] 
        \colhead{GeV/n} & \colhead{intensity} &
        \colhead{GeV/n} & \colhead{intensity} &
        \colhead{GeV/n} & \colhead{intensity} &
        \colhead{GeV/n} & \colhead{intensity} &
        \colhead{GeV/n} & \colhead{intensity}  
    }
\startdata
1.000e-03 & 1.354e+03 & 4.725e-02 & 2.222e+03 & 2.233e+00 & 3.918e+01 & 1.055e+02 & 2.522e-03 & 4.984e+03 & 1.365e-07\\
1.101e-03 & 1.489e+03 & 5.203e-02 & 2.170e+03 & 2.458e+00 & 3.228e+01 & 1.162e+02 & 1.948e-03 & 5.489e+03 & 1.069e-07\\
1.213e-03 & 1.530e+03 & 5.730e-02 & 2.114e+03 & 2.707e+00 & 2.650e+01 & 1.279e+02 & 1.506e-03 & 6.044e+03 & 8.370e-08\\
1.335e-03 & 1.567e+03 & 6.309e-02 & 2.055e+03 & 2.981e+00 & 2.168e+01 & 1.409e+02 & 1.166e-03 & 6.656e+03 & 6.548e-08\\
1.470e-03 & 1.605e+03 & 6.948e-02 & 1.992e+03 & 3.283e+00 & 1.768e+01 & 1.551e+02 & 9.038e-04 & 7.329e+03 & 5.119e-08\\
1.619e-03 & 1.643e+03 & 7.651e-02 & 1.927e+03 & 3.615e+00 & 1.436e+01 & 1.708e+02 & 7.014e-04 & 8.071e+03 & 3.999e-08\\
1.783e-03 & 1.683e+03 & 8.425e-02 & 1.859e+03 & 3.981e+00 & 1.163e+01 & 1.881e+02 & 5.450e-04 & 8.887e+03 & 3.122e-08\\
1.963e-03 & 1.723e+03 & 9.277e-02 & 1.788e+03 & 4.384e+00 & 9.383e+00 & 2.071e+02 & 4.240e-04 & 9.786e+03 & 2.434e-08\\
\enddata
    \tablecomments{Differential Intensity units: (m$^2$ s sr GeV/n)$^{-1}$.}
\end{deluxetable*}

\begin{deluxetable*}{cccccccccc}
\tabletypesize{\footnotesize}
    \tablecolumns{10}
    \tablewidth{0mm}
    \tablecaption{$^4$He LIS, standard model \label{Tbl-He4LIS-Rigi-default}}
    \tablehead{
        \colhead{Rigidity} & \colhead{Differential} &
        \colhead{Rigidity} & \colhead{Differential} &
        \colhead{Rigidity} & \colhead{Differential} &
        \colhead{Rigidity} & \colhead{Differential} &
        \colhead{Rigidity} & \colhead{Differential} \\ [-2ex] 
        \colhead{GV} & \colhead{intensity} &
        \colhead{GV} & \colhead{intensity} &
        \colhead{GV} & \colhead{intensity} &
        \colhead{GV} & \colhead{intensity} &
        \colhead{GV} & \colhead{intensity}  
    }
\startdata
8.635e-02 & 3.135e+01 & 6.009e-01 & 3.410e+02 & 6.048e+00 & 1.872e+01 & 2.128e+02 & 1.261e-03 & 9.970e+03 & 6.827e-08\\
9.061e-02 & 3.618e+01 & 6.313e-01 & 3.482e+02 & 6.519e+00 & 1.552e+01 & 2.342e+02 & 9.740e-04 & 1.098e+04 & 5.346e-08\\
9.509e-02 & 3.899e+01 & 6.634e-01 & 3.545e+02 & 7.035e+00 & 1.281e+01 & 2.577e+02 & 7.532e-04 & 1.209e+04 & 4.185e-08\\
9.979e-02 & 4.191e+01 & 6.972e-01 & 3.601e+02 & 7.600e+00 & 1.053e+01 & 2.836e+02 & 5.831e-04 & 1.331e+04 & 3.274e-08\\
1.047e-01 & 4.503e+01 & 7.328e-01 & 3.646e+02 & 8.220e+00 & 8.620e+00 & 3.121e+02 & 4.519e-04 & 1.466e+04 & 2.560e-08\\
1.099e-01 & 4.838e+01 & 7.704e-01 & 3.682e+02 & 8.900e+00 & 7.028e+00 & 3.435e+02 & 3.507e-04 & 1.614e+04 & 2.000e-08\\
1.153e-01 & 5.198e+01 & 8.101e-01 & 3.706e+02 & 9.646e+00 & 5.708e+00 & 3.780e+02 & 2.725e-04 & 1.778e+04 & 1.561e-08\\
1.210e-01 & 5.584e+01 & 8.519e-01 & 3.718e+02 & 1.047e+01 & 4.619e+00 & 4.161e+02 & 2.120e-04 & 1.957e+04 & 1.217e-08\\
\enddata
    \tablecomments{Differential Intensity units: (m$^2$ s sr GV)$^{-1}$.}
\end{deluxetable*}

\begin{deluxetable*}{cccccccccc}
\tabletypesize{\footnotesize}
    \tablecolumns{10}
    \tablewidth{0mm}
    \tablecaption{Total He LIS, standard model \label{Tbl-HeliumLIS-EKin-default}}
    \tablehead{
        \colhead{$E_{\rm kin}$} & \colhead{Differential} &
        \colhead{$E_{\rm kin}$} & \colhead{Differential} &
        \colhead{$E_{\rm kin}$} & \colhead{Differential} &
        \colhead{$E_{\rm kin}$} & \colhead{Differential} &
        \colhead{$E_{\rm kin}$} & \colhead{Differential} \\ [-2ex] 
        \colhead{GeV/n} & \colhead{intensity} &
        \colhead{GeV/n} & \colhead{intensity} &
        \colhead{GeV/n} & \colhead{intensity} &
        \colhead{GeV/n} & \colhead{intensity} &
        \colhead{GeV/n} & \colhead{intensity}  
    }
\startdata
1.000e-03 & 1.360e+03 & 4.725e-02 & 2.374e+03 & 2.233e+00 & 4.776e+01 & 1.055e+02 & 2.662e-03 & 4.984e+03 & 1.383e-07\\
1.101e-03 & 1.496e+03 & 5.203e-02 & 2.329e+03 & 2.458e+00 & 3.921e+01 & 1.162e+02 & 2.052e-03 & 5.489e+03 & 1.082e-07\\
1.213e-03 & 1.537e+03 & 5.730e-02 & 2.280e+03 & 2.707e+00 & 3.206e+01 & 1.279e+02 & 1.584e-03 & 6.044e+03 & 8.468e-08\\
1.335e-03 & 1.575e+03 & 6.309e-02 & 2.226e+03 & 2.981e+00 & 2.611e+01 & 1.409e+02 & 1.224e-03 & 6.656e+03 & 6.622e-08\\
1.470e-03 & 1.613e+03 & 6.948e-02 & 2.168e+03 & 3.283e+00 & 2.119e+01 & 1.551e+02 & 9.473e-04 & 7.329e+03 & 5.175e-08\\
1.619e-03 & 1.652e+03 & 7.651e-02 & 2.107e+03 & 3.615e+00 & 1.713e+01 & 1.708e+02 & 7.339e-04 & 8.071e+03 & 4.042e-08\\
1.783e-03 & 1.692e+03 & 8.425e-02 & 2.041e+03 & 3.981e+00 & 1.380e+01 & 1.881e+02 & 5.694e-04 & 8.887e+03 & 3.153e-08\\
1.963e-03 & 1.733e+03 & 9.277e-02 & 1.972e+03 & 4.384e+00 & 1.108e+01 & 2.071e+02 & 4.423e-04 & 9.786e+03 & 2.458e-08\\
\enddata
    \tablecomments{Differential Intensity units: (m$^2$ s sr GeV/n)$^{-1}$.}
\end{deluxetable*}

\begin{deluxetable*}{cccccccccc}
\tabletypesize{\footnotesize}
    \tablecolumns{10}
    \tablewidth{0mm}
    \tablecaption{Total He LIS, standard model \label{Tbl-HeliumLIS-Rigi-default}}
    \tablehead{
        \colhead{Rigidity} & \colhead{Differential} &
        \colhead{Rigidity} & \colhead{Differential} &
        \colhead{Rigidity} & \colhead{Differential} &
        \colhead{Rigidity} & \colhead{Differential} &
        \colhead{Rigidity} & \colhead{Differential} \\ [-2ex] 
        \colhead{GV} & \colhead{intensity} &
        \colhead{GV} & \colhead{intensity} &
        \colhead{GV} & \colhead{intensity} &
        \colhead{GV} & \colhead{intensity} &
        \colhead{GV} & \colhead{intensity}  
    }
\startdata
8.635e-02 & 3.175e+01 & 6.009e-01 & 3.889e+02 & 6.048e+00 & 2.115e+01 & 2.128e+02 & 1.299e-03 & 9.970e+03 & 6.877e-08\\
9.061e-02 & 3.663e+01 & 6.313e-01 & 3.986e+02 & 6.519e+00 & 1.747e+01 & 2.342e+02 & 1.003e-03 & 1.098e+04 & 5.384e-08\\
9.509e-02 & 3.951e+01 & 6.634e-01 & 4.072e+02 & 7.035e+00 & 1.437e+01 & 2.577e+02 & 7.748e-04 & 1.209e+04 & 4.213e-08\\
9.979e-02 & 4.249e+01 & 6.972e-01 & 4.148e+02 & 7.600e+00 & 1.177e+01 & 2.836e+02 & 5.993e-04 & 1.331e+04 & 3.295e-08\\
1.047e-01 & 4.570e+01 & 7.328e-01 & 4.211e+02 & 8.220e+00 & 9.596e+00 & 3.121e+02 & 4.641e-04 & 1.466e+04 & 2.576e-08\\
1.099e-01 & 4.914e+01 & 7.704e-01 & 4.260e+02 & 8.900e+00 & 7.795e+00 & 3.435e+02 & 3.598e-04 & 1.614e+04 & 2.012e-08\\
1.153e-01 & 5.285e+01 & 8.101e-01 & 4.294e+02 & 9.646e+00 & 6.308e+00 & 3.780e+02 & 2.794e-04 & 1.778e+04 & 1.570e-08\\
1.210e-01 & 5.683e+01 & 8.519e-01 & 4.311e+02 & 1.047e+01 & 5.086e+00 & 4.161e+02 & 2.172e-04 & 1.957e+04 & 1.224e-08\\
\enddata
    \tablecomments{Differential Intensity units: (m$^2$ s sr GV)$^{-1}$.}
\end{deluxetable*}

\begin{deluxetable*}{cccccccccc}
\tabletypesize{\footnotesize}
    \tablecolumns{10}
    \tablewidth{0mm}
    \tablecaption{$^3$He LIS, model with primary $^3$He \label{Tbl-He3LIS-EKin-prim}}
    \tablehead{
        \colhead{$E_{\rm kin}$} & \colhead{Differential} &
        \colhead{$E_{\rm kin}$} & \colhead{Differential} &
        \colhead{$E_{\rm kin}$} & \colhead{Differential} &
        \colhead{$E_{\rm kin}$} & \colhead{Differential} &
        \colhead{$E_{\rm kin}$} & \colhead{Differential} \\ [-2ex] 
        \colhead{GeV/n} & \colhead{intensity} &
        \colhead{GeV/n} & \colhead{intensity} &
        \colhead{GeV/n} & \colhead{intensity} &
        \colhead{GeV/n} & \colhead{intensity} &
        \colhead{GeV/n} & \colhead{intensity}  
    }
\startdata
1.000e-03 & 5.731e+00 & 4.725e-02 & 1.530e+02 & 2.233e+00 & 8.776e+00 & 1.055e+02 & 2.102e-04 & 4.984e+03 & 3.220e-09\\
1.101e-03 & 6.565e+00 & 5.203e-02 & 1.601e+02 & 2.458e+00 & 7.113e+00 & 1.162e+02 & 1.581e-04 & 5.489e+03 & 2.441e-09\\
1.213e-03 & 7.122e+00 & 5.730e-02 & 1.664e+02 & 2.707e+00 & 5.728e+00 & 1.279e+02 & 1.190e-04 & 6.044e+03 & 1.851e-09\\
1.335e-03 & 7.713e+00 & 6.309e-02 & 1.719e+02 & 2.981e+00 & 4.590e+00 & 1.409e+02 & 8.964e-05 & 6.656e+03 & 1.403e-09\\
1.470e-03 & 8.355e+00 & 6.948e-02 & 1.766e+02 & 3.283e+00 & 3.660e+00 & 1.551e+02 & 6.758e-05 & 7.329e+03 & 1.063e-09\\
1.619e-03 & 9.052e+00 & 7.651e-02 & 1.803e+02 & 3.615e+00 & 2.905e+00 & 1.708e+02 & 5.099e-05 & 8.071e+03 & 8.051e-10\\
1.783e-03 & 9.810e+00 & 8.425e-02 & 1.829e+02 & 3.981e+00 & 2.297e+00 & 1.881e+02 & 3.851e-05 & 8.887e+03 & 6.097e-10\\
1.963e-03 & 1.063e+01 & 9.277e-02 & 1.845e+02 & 4.384e+00 & 1.810e+00 & 2.071e+02 & 2.910e-05 & 9.786e+03 & 4.615e-10\\
\enddata
    \tablecomments{Differential Intensity units: (m$^2$ s sr GeV/n)$^{-1}$.}
\end{deluxetable*}

\begin{deluxetable*}{cccccccccc}
\tabletypesize{\footnotesize}
    \tablecolumns{10}
    \tablewidth{0mm}
    \tablecaption{$^3$He LIS, model with primary $^3$He \label{Tbl-He3LIS-Rigi-prim}}
    \tablehead{
        \colhead{Rigidity} & \colhead{Differential} &
        \colhead{Rigidity} & \colhead{Differential} &
        \colhead{Rigidity} & \colhead{Differential} &
        \colhead{Rigidity} & \colhead{Differential} &
        \colhead{Rigidity} & \colhead{Differential} \\ [-2ex] 
        \colhead{GV} & \colhead{intensity} &
        \colhead{GV} & \colhead{intensity} &
        \colhead{GV} & \colhead{intensity} &
        \colhead{GV} & \colhead{intensity} &
        \colhead{GV} & \colhead{intensity}  
    }
\startdata
6.476e-02 & 1.769e-01 & 4.506e-01 & 3.131e+01 & 4.536e+00 & 5.592e+00 & 1.596e+02 & 1.401e-04 & 7.478e+03 & 2.146e-09\\
6.796e-02 & 2.126e-01 & 4.735e-01 & 3.425e+01 & 4.889e+00 & 4.560e+00 & 1.756e+02 & 1.054e-04 & 8.234e+03 & 1.627e-09\\
7.132e-02 & 2.420e-01 & 4.975e-01 & 3.721e+01 & 5.276e+00 & 3.692e+00 & 1.933e+02 & 7.933e-05 & 9.067e+03 & 1.234e-09\\
7.484e-02 & 2.750e-01 & 5.229e-01 & 4.018e+01 & 5.700e+00 & 2.972e+00 & 2.127e+02 & 5.976e-05 & 9.985e+03 & 9.351e-10\\
7.854e-02 & 3.126e-01 & 5.496e-01 & 4.310e+01 & 6.165e+00 & 2.380e+00 & 2.341e+02 & 4.505e-05 & 1.099e+04 & 7.086e-10\\
8.242e-02 & 3.554e-01 & 5.778e-01 & 4.593e+01 & 6.675e+00 & 1.896e+00 & 2.576e+02 & 3.399e-05 & 1.211e+04 & 5.367e-10\\
8.649e-02 & 4.041e-01 & 6.075e-01 & 4.863e+01 & 7.235e+00 & 1.503e+00 & 2.835e+02 & 2.567e-05 & 1.333e+04 & 4.065e-10\\
9.077e-02 & 4.596e-01 & 6.389e-01 & 5.116e+01 & 7.849e+00 & 1.188e+00 & 3.121e+02 & 1.940e-05 & 1.468e+04 & 3.077e-10\\
\enddata
    \tablecomments{Differential Intensity units: (m$^2$ s sr GV)$^{-1}$.}
\end{deluxetable*}

\begin{deluxetable*}{cccccccccc}
\tabletypesize{\footnotesize}
    \tablecolumns{10}
    \tablewidth{0mm}
    \tablecaption{Total He LIS, model with primary $^3$He \label{Tbl-HeliumLIS-EKin-prim}}
    \tablehead{
        \colhead{$E_{\rm kin}$} & \colhead{Differential} &
        \colhead{$E_{\rm kin}$} & \colhead{Differential} &
        \colhead{$E_{\rm kin}$} & \colhead{Differential} &
        \colhead{$E_{\rm kin}$} & \colhead{Differential} &
        \colhead{$E_{\rm kin}$} & \colhead{Differential} \\ [-2ex] 
        \colhead{GeV/n} & \colhead{intensity} &
        \colhead{GeV/n} & \colhead{intensity} &
        \colhead{GeV/n} & \colhead{intensity} &
        \colhead{GeV/n} & \colhead{intensity} &
        \colhead{GeV/n} & \colhead{intensity}  
    }
\startdata
1.000e-03 & 1.360e+03 & 4.725e-02 & 2.375e+03 & 2.233e+00 & 4.796e+01 & 1.055e+02 & 2.732e-03 & 4.984e+03 & 1.398e-07\\
1.101e-03 & 1.496e+03 & 5.203e-02 & 2.330e+03 & 2.458e+00 & 3.940e+01 & 1.162e+02 & 2.106e-03 & 5.489e+03 & 1.094e-07\\
1.213e-03 & 1.537e+03 & 5.730e-02 & 2.280e+03 & 2.707e+00 & 3.223e+01 & 1.279e+02 & 1.625e-03 & 6.044e+03 & 8.555e-08\\
1.335e-03 & 1.575e+03 & 6.309e-02 & 2.227e+03 & 2.981e+00 & 2.627e+01 & 1.409e+02 & 1.256e-03 & 6.656e+03 & 6.688e-08\\
1.470e-03 & 1.613e+03 & 6.948e-02 & 2.169e+03 & 3.283e+00 & 2.134e+01 & 1.551e+02 & 9.714e-04 & 7.329e+03 & 5.226e-08\\
1.619e-03 & 1.652e+03 & 7.651e-02 & 2.107e+03 & 3.615e+00 & 1.727e+01 & 1.708e+02 & 7.524e-04 & 8.071e+03 & 4.080e-08\\
1.783e-03 & 1.693e+03 & 8.425e-02 & 2.042e+03 & 3.981e+00 & 1.392e+01 & 1.881e+02 & 5.835e-04 & 8.887e+03 & 3.183e-08\\
1.963e-03 & 1.734e+03 & 9.277e-02 & 1.973e+03 & 4.384e+00 & 1.119e+01 & 2.071e+02 & 4.531e-04 & 9.786e+03 & 2.480e-08\\
\enddata
    \tablecomments{Differential Intensity units: (m$^2$ s sr GeV/n)$^{-1}$.}
\end{deluxetable*}

\begin{deluxetable*}{cccccccccc}
\tabletypesize{\footnotesize}
    \tablecolumns{10}
    \tablewidth{0mm}
    \tablecaption{Total He LIS, model with primary $^3$He \label{Tbl-HeliumLIS-Rigi-prim}}
    \tablehead{
        \colhead{Rigidity} & \colhead{Differential} &
        \colhead{Rigidity} & \colhead{Differential} &
        \colhead{Rigidity} & \colhead{Differential} &
        \colhead{Rigidity} & \colhead{Differential} &
        \colhead{Rigidity} & \colhead{Differential} \\ [-2ex] 
        \colhead{GV} & \colhead{intensity} &
        \colhead{GV} & \colhead{intensity} &
        \colhead{GV} & \colhead{intensity} &
        \colhead{GV} & \colhead{intensity} &
        \colhead{GV} & \colhead{intensity}  
    }
\startdata
8.635e-02 & 3.175e+01 & 6.009e-01 & 3.890e+02 & 6.048e+00 & 2.125e+01 & 2.128e+02 & 1.320e-03 & 9.970e+03 & 6.921e-08\\
9.061e-02 & 3.663e+01 & 6.313e-01 & 3.987e+02 & 6.519e+00 & 1.756e+01 & 2.342e+02 & 1.019e-03 & 1.098e+04 & 5.418e-08\\
9.509e-02 & 3.951e+01 & 6.634e-01 & 4.074e+02 & 7.035e+00 & 1.445e+01 & 2.577e+02 & 7.871e-04 & 1.209e+04 & 4.239e-08\\
9.979e-02 & 4.250e+01 & 6.972e-01 & 4.150e+02 & 7.600e+00 & 1.185e+01 & 2.836e+02 & 6.087e-04 & 1.331e+04 & 3.315e-08\\
1.047e-01 & 4.570e+01 & 7.328e-01 & 4.213e+02 & 8.220e+00 & 9.669e+00 & 3.121e+02 & 4.713e-04 & 1.466e+04 & 2.591e-08\\
1.099e-01 & 4.915e+01 & 7.704e-01 & 4.262e+02 & 8.900e+00 & 7.862e+00 & 3.435e+02 & 3.654e-04 & 1.614e+04 & 2.023e-08\\
1.153e-01 & 5.286e+01 & 8.101e-01 & 4.296e+02 & 9.646e+00 & 6.369e+00 & 3.780e+02 & 2.836e-04 & 1.778e+04 & 1.579e-08\\
1.210e-01 & 5.684e+01 & 8.519e-01 & 4.313e+02 & 1.047e+01 & 5.140e+00 & 4.161e+02 & 2.204e-04 & 1.957e+04 & 1.230e-08\\
\enddata
    \tablecomments{Differential Intensity units: (m$^2$ s sr GV)$^{-1}$.}
\end{deluxetable*}

\begin{deluxetable*}{cccccccccc}
\tabletypesize{\footnotesize}
    \tablecolumns{10}
    \tablewidth{0mm}
    \tablecaption{$^3$He LIS, alternative model \label{Tbl-He3LIS-EKin-alternative}}
    \tablehead{
        \colhead{$E_{\rm kin}$} & \colhead{Differential} &
        \colhead{$E_{\rm kin}$} & \colhead{Differential} &
        \colhead{$E_{\rm kin}$} & \colhead{Differential} &
        \colhead{$E_{\rm kin}$} & \colhead{Differential} &
        \colhead{$E_{\rm kin}$} & \colhead{Differential} \\ [-2ex] 
        \colhead{GeV/n} & \colhead{intensity} &
        \colhead{GeV/n} & \colhead{intensity} &
        \colhead{GeV/n} & \colhead{intensity} &
        \colhead{GeV/n} & \colhead{intensity} &
        \colhead{GeV/n} & \colhead{intensity}  
    }
\startdata
1.000e-03 & 4.459e+00 & 1.226e-02 & 4.265e+01 & 1.502e-01 & 1.525e+02 & 1.841e+00 & 1.281e+01 & 2.257e+01 & 2.074e-02\\
1.101e-03 & 5.389e+00 & 1.350e-02 & 4.650e+01 & 1.654e-01 & 1.485e+02 & 2.027e+00 & 1.061e+01 & 2.485e+01 & 1.572e-02\\
1.213e-03 & 5.880e+00 & 1.486e-02 & 5.073e+01 & 1.822e-01 & 1.436e+02 & 2.233e+00 & 8.712e+00 & 2.736e+01 & 1.191e-02\\
1.335e-03 & 6.372e+00 & 1.637e-02 & 5.539e+01 & 2.006e-01 & 1.380e+02 & 2.458e+00 & 7.098e+00 & 3.013e+01 & 9.018e-03\\
1.470e-03 & 6.903e+00 & 1.802e-02 & 6.052e+01 & 2.209e-01 & 1.318e+02 & 2.707e+00 & 5.743e+00 & 3.318e+01 & 6.826e-03\\
1.619e-03 & 7.480e+00 & 1.985e-02 & 6.622e+01 & 2.432e-01 & 1.253e+02 & 2.981e+00 & 4.620e+00 & 3.654e+01 & 5.167e-03\\
1.783e-03 & 8.107e+00 & 2.185e-02 & 7.258e+01 & 2.678e-01 & 1.186e+02 & 3.283e+00 & 3.695e+00 & 4.023e+01 & 3.910e-03\\
1.963e-03 & 8.788e+00 & 2.406e-02 & 7.972e+01 & 2.949e-01 & 1.118e+02 & 3.615e+00 & 2.940e+00 & 4.431e+01 & 2.959e-03\\
\enddata
    \tablecomments{Differential Intensity units: (m$^2$ s sr GeV/n)$^{-1}$.}
\end{deluxetable*}

\begin{deluxetable*}{cccccccccc}
\tabletypesize{\footnotesize}
    \tablecolumns{10}
    \tablewidth{0mm}
    \tablecaption{$^3$He LIS, alternative model \label{Tbl-He3LIS-Rigi-alternative}}
    \tablehead{
        \colhead{Rigidity} & \colhead{Differential} &
        \colhead{Rigidity} & \colhead{Differential} &
        \colhead{Rigidity} & \colhead{Differential} &
        \colhead{Rigidity} & \colhead{Differential} &
        \colhead{Rigidity} & \colhead{Differential} \\ [-2ex] 
        \colhead{GV} & \colhead{intensity} &
        \colhead{GV} & \colhead{intensity} &
        \colhead{GV} & \colhead{intensity} &
        \colhead{GV} & \colhead{intensity} &
        \colhead{GV} & \colhead{intensity}  
    }
\startdata
6.476e-02 & 1.376e-01 & 2.274e-01 & 4.568e+00 & 8.249e-01 & 5.168e+01 & 3.917e+00 & 8.042e+00 & 3.522e+01 & 1.382e-02\\
6.796e-02 & 1.745e-01 & 2.387e-01 & 5.221e+00 & 8.689e-01 & 5.227e+01 & 4.213e+00 & 6.713e+00 & 3.865e+01 & 1.047e-02\\
7.132e-02 & 1.998e-01 & 2.506e-01 & 5.970e+00 & 9.155e-01 & 5.247e+01 & 4.536e+00 & 5.551e+00 & 4.242e+01 & 7.935e-03\\
7.484e-02 & 2.272e-01 & 2.631e-01 & 6.832e+00 & 9.651e-01 & 5.228e+01 & 4.889e+00 & 4.550e+00 & 4.657e+01 & 6.009e-03\\
7.854e-02 & 2.583e-01 & 2.762e-01 & 7.824e+00 & 1.018e+00 & 5.174e+01 & 5.276e+00 & 3.701e+00 & 5.115e+01 & 4.549e-03\\
8.242e-02 & 2.936e-01 & 2.900e-01 & 8.971e+00 & 1.074e+00 & 5.090e+01 & 5.700e+00 & 2.991e+00 & 5.619e+01 & 3.443e-03\\
8.649e-02 & 3.339e-01 & 3.044e-01 & 1.030e+01 & 1.133e+00 & 4.981e+01 & 6.165e+00 & 2.403e+00 & 6.173e+01 & 2.606e-03\\
9.077e-02 & 3.798e-01 & 3.196e-01 & 1.185e+01 & 1.197e+00 & 4.849e+01 & 6.675e+00 & 1.918e+00 & 6.784e+01 & 1.972e-03\\
\enddata
    \tablecomments{Differential Intensity units: (m$^2$ s sr GV)$^{-1}$.}
\end{deluxetable*}

\begin{deluxetable*}{cccccccccc}
\tabletypesize{\footnotesize}
    \tablecolumns{10}
    \tablewidth{0mm}
    \tablecaption{$^4$He LIS, alternative model \label{Tbl-He4LIS-EKin-alternative}}
    \tablehead{
        \colhead{$E_{\rm kin}$} & \colhead{Differential} &
        \colhead{$E_{\rm kin}$} & \colhead{Differential} &
        \colhead{$E_{\rm kin}$} & \colhead{Differential} &
        \colhead{$E_{\rm kin}$} & \colhead{Differential} &
        \colhead{$E_{\rm kin}$} & \colhead{Differential} \\ [-2ex] 
        \colhead{GeV/n} & \colhead{intensity} &
        \colhead{GeV/n} & \colhead{intensity} &
        \colhead{GeV/n} & \colhead{intensity} &
        \colhead{GeV/n} & \colhead{intensity} &
        \colhead{GeV/n} & \colhead{intensity}  
    }
\startdata
1.000e-03 & 1.029e+03 & 1.226e-02 & 2.167e+03 & 1.502e-01 & 1.306e+03 & 1.841e+00 & 5.689e+01 & 2.257e+01 & 1.591e-01\\
1.101e-03 & 1.222e+03 & 1.350e-02 & 2.187e+03 & 1.654e-01 & 1.236e+03 & 2.027e+00 & 4.706e+01 & 2.485e+01 & 1.237e-01\\
1.213e-03 & 1.272e+03 & 1.486e-02 & 2.204e+03 & 1.822e-01 & 1.166e+03 & 2.233e+00 & 3.879e+01 & 2.736e+01 & 9.607e-02\\
1.335e-03 & 1.308e+03 & 1.637e-02 & 2.215e+03 & 2.006e-01 & 1.096e+03 & 2.458e+00 & 3.187e+01 & 3.013e+01 & 7.458e-02\\
1.470e-03 & 1.343e+03 & 1.802e-02 & 2.222e+03 & 2.209e-01 & 1.025e+03 & 2.707e+00 & 2.609e+01 & 3.318e+01 & 5.787e-02\\
1.619e-03 & 1.379e+03 & 1.985e-02 & 2.224e+03 & 2.432e-01 & 9.550e+02 & 2.981e+00 & 2.128e+01 & 3.654e+01 & 4.489e-02\\
1.783e-03 & 1.416e+03 & 2.185e-02 & 2.220e+03 & 2.678e-01 & 8.854e+02 & 3.283e+00 & 1.730e+01 & 4.023e+01 & 3.481e-02\\
1.963e-03 & 1.454e+03 & 2.406e-02 & 2.210e+03 & 2.949e-01 & 8.168e+02 & 3.615e+00 & 1.402e+01 & 4.431e+01 & 2.700e-02\\
\enddata
    \tablecomments{Differential Intensity units: (m$^2$ s sr GeV/n)$^{-1}$.}
\end{deluxetable*}

\begin{deluxetable*}{cccccccccc}
\tabletypesize{\footnotesize}
    \tablecolumns{10}
    \tablewidth{0mm}
    \tablecaption{$^4$He LIS, alternative model \label{Tbl-He4LIS-Rigi-alternative}}
    \tablehead{
        \colhead{Rigidity} & \colhead{Differential} &
        \colhead{Rigidity} & \colhead{Differential} &
        \colhead{Rigidity} & \colhead{Differential} &
        \colhead{Rigidity} & \colhead{Differential} &
        \colhead{Rigidity} & \colhead{Differential} \\ [-2ex] 
        \colhead{GV} & \colhead{intensity} &
        \colhead{GV} & \colhead{intensity} &
        \colhead{GV} & \colhead{intensity} &
        \colhead{GV} & \colhead{intensity} &
        \colhead{GV} & \colhead{intensity}  
    }
\startdata
8.635e-02 & 2.383e+01 & 3.032e-01 & 1.740e+02 & 1.100e+00 & 3.320e+02 & 5.223e+00 & 2.679e+01 & 4.696e+01 & 7.951e-02\\
9.061e-02 & 2.967e+01 & 3.183e-01 & 1.842e+02 & 1.159e+00 & 3.264e+02 & 5.617e+00 & 2.233e+01 & 5.153e+01 & 6.180e-02\\
9.509e-02 & 3.241e+01 & 3.341e-01 & 1.945e+02 & 1.221e+00 & 3.196e+02 & 6.048e+00 & 1.854e+01 & 5.656e+01 & 4.801e-02\\
9.979e-02 & 3.497e+01 & 3.508e-01 & 2.049e+02 & 1.287e+00 & 3.113e+02 & 6.519e+00 & 1.532e+01 & 6.210e+01 & 3.727e-02\\
1.047e-01 & 3.768e+01 & 3.682e-01 & 2.155e+02 & 1.357e+00 & 3.018e+02 & 7.035e+00 & 1.261e+01 & 6.820e+01 & 2.892e-02\\
1.099e-01 & 4.060e+01 & 3.866e-01 & 2.259e+02 & 1.432e+00 & 2.909e+02 & 7.600e+00 & 1.033e+01 & 7.492e+01 & 2.244e-02\\
1.153e-01 & 4.374e+01 & 4.059e-01 & 2.363e+02 & 1.511e+00 & 2.789e+02 & 8.220e+00 & 8.437e+00 & 8.231e+01 & 1.740e-02\\
1.210e-01 & 4.712e+01 & 4.262e-01 & 2.465e+02 & 1.596e+00 & 2.657e+02 & 8.900e+00 & 6.863e+00 & 9.046e+01 & 1.350e-02\\
\enddata
    \tablecomments{Differential Intensity units: (m$^2$ s sr GV)$^{-1}$.}
\end{deluxetable*}

\begin{deluxetable*}{cccccccccc}
\tabletypesize{\footnotesize}
    \tablecolumns{10}
    \tablewidth{0mm}
    \tablecaption{Total He LIS, alternative model \label{Tbl-HeliumLIS-EKin-alternative}}
    \tablehead{
        \colhead{$E_{\rm kin}$} & \colhead{Differential} &
        \colhead{$E_{\rm kin}$} & \colhead{Differential} &
        \colhead{$E_{\rm kin}$} & \colhead{Differential} &
        \colhead{$E_{\rm kin}$} & \colhead{Differential} &
        \colhead{$E_{\rm kin}$} & \colhead{Differential} \\ [-2ex] 
        \colhead{GeV/n} & \colhead{intensity} &
        \colhead{GeV/n} & \colhead{intensity} &
        \colhead{GeV/n} & \colhead{intensity} &
        \colhead{GeV/n} & \colhead{intensity} &
        \colhead{GeV/n} & \colhead{intensity}  
    }
\startdata
1.000e-03 & 1.034e+03 & 1.226e-02 & 2.210e+03 & 1.502e-01 & 1.458e+03 & 1.841e+00 & 6.969e+01 & 2.257e+01 & 1.799e-01\\
1.101e-03 & 1.227e+03 & 1.350e-02 & 2.234e+03 & 1.654e-01 & 1.385e+03 & 2.027e+00 & 5.767e+01 & 2.485e+01 & 1.394e-01\\
1.213e-03 & 1.277e+03 & 1.486e-02 & 2.254e+03 & 1.822e-01 & 1.310e+03 & 2.233e+00 & 4.751e+01 & 2.736e+01 & 1.080e-01\\
1.335e-03 & 1.314e+03 & 1.637e-02 & 2.271e+03 & 2.006e-01 & 1.234e+03 & 2.458e+00 & 3.896e+01 & 3.013e+01 & 8.359e-02\\
1.470e-03 & 1.350e+03 & 1.802e-02 & 2.283e+03 & 2.209e-01 & 1.157e+03 & 2.707e+00 & 3.183e+01 & 3.318e+01 & 6.469e-02\\
1.619e-03 & 1.386e+03 & 1.985e-02 & 2.290e+03 & 2.432e-01 & 1.080e+03 & 2.981e+00 & 2.590e+01 & 3.654e+01 & 5.006e-02\\
1.783e-03 & 1.424e+03 & 2.185e-02 & 2.293e+03 & 2.678e-01 & 1.004e+03 & 3.283e+00 & 2.100e+01 & 4.023e+01 & 3.872e-02\\
1.963e-03 & 1.463e+03 & 2.406e-02 & 2.290e+03 & 2.949e-01 & 9.286e+02 & 3.615e+00 & 1.696e+01 & 4.431e+01 & 2.996e-02\\
\enddata
    \tablecomments{Differential Intensity units: (m$^2$ s sr GeV/n)$^{-1}$.}
\end{deluxetable*}

\begin{deluxetable*}{cccccccccc}
\tabletypesize{\footnotesize}
    \tablecolumns{10}
    \tablewidth{0mm}
    \tablecaption{Total He LIS, alternative model \label{Tbl-HeliumLIS-Rigi-alternative}}
    \tablehead{
        \colhead{Rigidity} & \colhead{Differential} &
        \colhead{Rigidity} & \colhead{Differential} &
        \colhead{Rigidity} & \colhead{Differential} &
        \colhead{Rigidity} & \colhead{Differential} &
        \colhead{Rigidity} & \colhead{Differential} \\ [-2ex] 
        \colhead{GV} & \colhead{intensity} &
        \colhead{GV} & \colhead{intensity} &
        \colhead{GV} & \colhead{intensity} &
        \colhead{GV} & \colhead{intensity} &
        \colhead{GV} & \colhead{intensity}  
    }
\startdata
8.635e-02 & 2.416e+01 & 2.889e-01 & 1.730e+02 & 9.922e-01 & 3.910e+02 & 4.228e+00 & 5.165e+01 & 3.250e+01 & 2.335e-01\\
9.061e-02 & 3.005e+01 & 3.032e-01 & 1.842e+02 & 1.045e+00 & 3.875e+02 & 4.532e+00 & 4.360e+01 & 3.561e+01 & 1.818e-01\\
9.509e-02 & 3.284e+01 & 3.183e-01 & 1.959e+02 & 1.100e+00 & 3.824e+02 & 4.862e+00 & 3.662e+01 & 3.904e+01 & 1.415e-01\\
9.979e-02 & 3.546e+01 & 3.341e-01 & 2.078e+02 & 1.159e+00 & 3.757e+02 & 5.223e+00 & 3.061e+01 & 4.281e+01 & 1.100e-01\\
1.047e-01 & 3.824e+01 & 3.508e-01 & 2.200e+02 & 1.221e+00 & 3.675e+02 & 5.617e+00 & 2.546e+01 & 4.696e+01 & 8.540e-02\\
1.099e-01 & 4.124e+01 & 3.682e-01 & 2.324e+02 & 1.287e+00 & 3.578e+02 & 6.048e+00 & 2.109e+01 & 5.153e+01 & 6.627e-02\\
1.153e-01 & 4.446e+01 & 3.866e-01 & 2.450e+02 & 1.357e+00 & 3.465e+02 & 6.519e+00 & 1.739e+01 & 5.656e+01 & 5.139e-02\\
1.210e-01 & 4.794e+01 & 4.059e-01 & 2.576e+02 & 1.432e+00 & 3.339e+02 & 7.035e+00 & 1.427e+01 & 6.210e+01 & 3.984e-02\\
\enddata
    \tablecomments{Differential Intensity units: (m$^2$ s sr GV)$^{-1}$.}
\end{deluxetable*}

\end{document}